 \documentclass[twocolumn]{aastex631}
\usepackage{amsmath}
\usepackage[caption=false]{subfig}

\begin{document}

\title{Foreground removal and angular power spectrum estimation of 21 cm signal using harmonic space ILC method}

\author{Albin Joseph}
\email{ajosep52@asu.edu}
\affiliation{Arizona State University, Tempe, AZ, USA}
\author{Rajib Saha}
\email{rajib@iiserb.ac.in}
\affiliation{Indian Institute of Science Education and Research Bhopal, MP, India}



\begin{abstract}

Mapping the distribution of neutral atomic hydrogen (HI) in the Universe through its 21 cm emission line provides a powerful cosmological probe to map the large-scale structures and shed light on various cosmological phenomena. The Baryon Acoustic Oscillations at low redshifts can potentially be probed by sensitive HI intensity mapping experiments and constrain the properties of dark energy. However, the 21 cm signal detection faces formidable challenges due to the dominance of various astrophysical foregrounds, which can be several orders of magnitude stronger. Our current work introduces a novel and model-independent Internal Linear Combination (ILC) method in harmonic space  using the principal components of the  21 cm signal  for accurate foreground removal and power spectrum estimation. We estimate the principal components  by incorporating prior knowledge of the theoretical 21 cm covariance matrix.  We test our methodology by detailed  simulations of  radio observations, incorporating synchrotron emission, free-free radiation, extragalactic point sources, and thermal noise.  We estimate the full sky 21 cm angular power spectrum after application of a mask on the full sky cleaned 21 cm signal by using the mode-mode coupling matrix. These full sky estimates of angular spectra can be directly used to measure the cosmological parameters. 
For the first time, we demonstrate the effectiveness of a foreground model-independent ILC method in harmonic space to reconstruct the 21 cm signal.

\end{abstract}



\section{Introduction}\label{intro}

Over the past years, advancements in 21 cm cosmology~\citep{2006PhR...433..181F,2010ARA&A..48..127M,2012RPPh...75h6901P} have markedly deepened our comprehension of the Universe. The 21 cm intensity mapping (IM) technique~\citep{2019BAAS...51c.101K,2001JApA...22..293B,2001JApA...22...21B,2004MNRAS.355.1339B,2006ApJ...653..815M,2008PhRvL.100i1303C,2015ApJ...803...21B} is a novel and promising tool for probing the intermediate stages of our cosmic timeline, which have not been systematically surveyed before. The 21 cm spectral line arises from the ``spin flip" transition of the neutral hydrogen (HI) atom. By observing the absorption or emission of the redshifted 21 cm line, one can study neutral hydrogen as a tracer for the distribution of matter in the Universe or as an indirect probe of other properties, such as ionization state or temperature~\citep{2006PhR...433..181F,2020PASP..132f2001L}. This approach can greatly improve our knowledge of the Universe by shedding light onto its Dark Ages~\citep{1990MNRAS.247..510S,2004PhRvL..92u1301L}, the Cosmic Dawn~\citep{2018Natur.555...71B,2018ApJ...858L..17F,2018ApJ...868...63E,2018arXiv180210094M,2019MNRAS.486.1763F}, and also its  post-reionization phases~\citep{2018ApJ...866..135V}.

The 21 cm intensity mapping involves measuring the integrated HI emission lines originating from numerous unresolved galaxies, allowing the tracing of large-scale structures (LSS) over cosmological distances. One of the main advantages of intensity mapping, compared to other optical galaxy surveys, is the ability to achieve a large sky volume within a relatively short observing time. The first detection of the 21 cm signal using the intensity mapping method was made by the Green Bank Telescope (GBT)~\citep{nature1}. By cross-correlating LSS from the DEEP2 optical galaxy redshift survey with 21 cm intensity maps at $z\sim 1$, the authors detected the 21 cm signal at a significance of approximately $4\sigma$. Subsequently, Masui~\citep{2013ApJ...763L..20M} improved the detection by cross-correlating the 21 cm data obtained with the GBT with optical data from the WiggleZ Dark Energy Survey~\citep{2010MNRAS.401.1429D}. Currently, several 21 cm intensity mapping experiments are operational or under development. Some experiments, like GBT, MeerKAT (The South African Square Kilometer Array Pathfinder)~\citep{2021MNRAS.505.3698W}, and BINGO (Baryon acoustic oscillations In Neutral Gas Observations)~\citep{2013MNRAS.434.1239B}, are designed as single-dish telescopes, while others, such as CHIME (Canadian Hydrogen Intensity Mapping Experiment)~\citep{2022ApJS..261...29C,2023ApJ...947...16A,2023arXiv230904404C}, GMRT (Giant Meterwave Radio Telescope)~\citep{2013MNRAS.433..639P,2020Natur.586..369C,2007MNRAS.376.1357L} and HIRAX (The Hydrogen Intensity and Real-time Analysis eXperiment)~\citep{2022JATIS...8a1019C}, are designed as interferometer arrays. The upcoming SKA (The Square Kilometer Array)~\citep{2015aska.confE...1K} project will enable intensity mapping surveys in single-dish mode within the redshift range of $0.35 < z < 3$ using SKA-MID and in interferometer mode for the redshift range of $3 < z < 5$ with SKA-LOW. Both types of experiments will need to address potential systematics, such as gain variations and correlated noise in frequency. Single-dish experiments require reliable receiver systems and precise calibration. In contrast, interferometers are inherently better at handling systematic issues compared to single-dish setups~\citep{1999ApJ...514...12W,2012rsri.confE..34D}. Nevertheless, it is noteworthy that existing interferometers are constrained by the small number of their shortest baselines, and consequently, they do not offer the essential surface brightness sensitivity~\citep{2015ApJ...803...21B}.

Detecting the 21 cm signal poses significant challenges due to the overwhelming presence of astrophysical foregrounds, which can be up to approximately $10^5$ times stronger than the 21 cm signal itself~\citep{2015PhRvD..91h3514S,2016MNRAS.456.2749O}. The major astrophysical foreground emissions that can corrupt the 21 cm signals are synchrotron emission, free-free emission, and the background emission of extragalactic point sources. The foregrounds, including synchrotron and free-free emission, exhibit smooth frequency spectra and are correlated over frequencies, whereas the 21 cm signal is only partially correlated over adjacent frequencies~\citep{2013ApJ...763L..20M,2015MNRAS.447..400A}. This disparity in spectral characteristics forms the basis for various methods employed in the literature to separate foregrounds from the 21 cm signal. One category of methods includes parametric or linear techniques, such as Karhunen–Loeve Decomposition~\citep{2014ApJ...781...57S} and the delay filter~\citep{2021MNRAS.500.5195E}. These methods rely on assuming a physical model for sky components based on their known properties but are limited by the assumption of a well-known instrumental response, leading to potential signal contamination due to foreground residuals~\citep{2022MNRAS.509.2048S}. Non-parametric or blind methods, on the other hand, perform component separation without assuming a specific model and are thus more versatile. Principal Component Analysis (PCA)\citep{2013ApJ...763L..20M,2013MNRAS.434L..46S,2015MNRAS.447..400A}, Independent Component Analysis (ICA)\citep{2015MNRAS.447..400A,2017MNRAS.464.4938W}, Generalised Morphological Component Analysis (GMCA)\citep{2020MNRAS.499..304C}, and Generalized Needlet Internal Linear Combination (GNILC)\citep{2011MNRAS.418..467R, 2016MNRAS.456.2749O} are examples of non-parametric methods.
A unified matrix-based framework for foreground subtraction and power spectrum estimation was developed in \cite{2011PhRvD..83j3006L}, allowing quantification of errors and biases that arise in the power spectrum due to foreground subtraction. Another significant approach combines Bayesian spatial regularization with GMCA for foreground removal, demonstrating successful recovery of the 21 cm power spectrum within specific k-space ranges while effectively managing noise in the reconstructed maps \citep{2015MNRAS.452.1587G}. Building on ICA methodology, the HIEMICA (HI Expectation-Maximization Independent Component Analysis) approach provides a Bayesian semi-blind technique that can jointly estimate the 3D power spectrum of the 21-cm signal and foreground properties without prior assumptions about foreground characteristics \citep{2016ApJS..222....3Z}.
The Gaussian Process Regression (GPR) method was developed to address both foreground removal and calibration-related contamination \citep{2018MNRAS.478.3640M}. However, subsequent theoretical analysis has shown that GPR-based foreground subtraction can distort window functions at low k modes where the EoR signal is strongest, highlighting the importance of careful consideration of statistical properties in foreground removal techniques \citep{2021MNRAS.501.1463K}. In~\cite{2019AJ....157....4Z} authors introduces  Robust Principal Component Analysis (RPCA) method  to subtract the foreground and extract the 21 cm signal, leveraging the sparsity and low-rank structure in the covariance matrix of the 21 cm signal and foregrounds. While Optimal Quadratic Estimator (OQE) methods are widely used for power spectrum estimation, they can suffer from reduced sensitivity and signal loss when covariance modeling is inaccurate. To address this, Bayesian approaches incorporating Gaussian constrained realisations and Gibbs sampling have been developed that can jointly estimate the 21 cm fluctuations, their power spectrum, and foreground emission while dynamically estimating covariance as part of the inference process \citep{2023ApJS..266...23K,2024MNRAS.535..793B}.
 Additionally, the hybrid foreground residual subtraction (HyFoReS) technique addresses beam-induced foreground contamination by cross-correlating linearly filtered data, demonstrating improved signal-to-noise ratios in 21 cm signal detection and enabling better recovery of large-scale modes \citep{2024arXiv240808949W}.

In the context of addressing the challenges posed by foregrounds in 21 cm signal detection, our current work focuses on studying a novel method for 21 cm foreground removal and power spectrum estimation. The first major challenge in extracting a cleaned 21 cm signal from the observations is the strong contaminations due to astrophysical foregrounds.  Such a problem would have been relatively simpler to solve  if the foreground   properties (e.g.,  spectral index  for synchrotron component)  were independent of the sky locations. In real life the synchrotron spectral index varies over sky positions. This variation makes the 21 cm reconstruction problem even more difficult since spectral variations introduce a larger number of foreground components (parameters) to be removed from the given observation~(\citep{2008PhRvD..78b3003S} and references therein).   We employ the harmonic space Internal Linear Combination (ILC) method, which effectively incorporates variation of foreground properties over different multipoles or angular scales.   Our results show that ILC is a powerful technique for component separation in intensity mapping experiments. The ILC method has been previously applied in pixel space as well as in harmonic space to successfully extract the CMB signal from foreground emissions~\citep{10.1111/j.1365-2966.2011.19770.x,PhysRevD.68.123523,SOURADEEP2006854,Eriksen_2008,2018ApJ...867...74S,2023MNRAS.520..976J}.

We structure our paper as follows. In Section~\ref{formalism}, we present the fundamental formalism of our work, outlining our algorithm for obtaining a cleaned 21 cm signal and its corresponding angular power spectrum. Section~\ref{sim} provides a detailed discussion of the input maps utilized in our analysis. Moving on to Section~\ref{results}, we present and discuss the methodology and the outcomes of our analysis. In Section~\ref{systematics}, we investigate the effects of various systematic effects on the foreground removal technique. Finally, we summarize our findings and draw conclusions in Section~\ref{conclusions}.

\section{Formalism}\label{formalism}
Observations of the radio sky capture a diverse array of signals originating from various cosmological (e.g., CMB and cosmological 21 cm signal) and astrophysical phenomena (e.g., Galactic foregrounds and extragalactic point sources), as well as instrumental noise (e.g., thermal noise). Component separation denotes the process of disentangling these signals by leveraging correlations across different frequencies, external constraints, and physical models of the different sources of emission~\citep{2007astro.ph..2198D}. This separation can be executed through parametric methods, such as Wiener filtering~\citep{1999MNRAS.302..663B}, Gibbs sampling approach~\citep{2008ApJ...676...10E,2004PhRvD..70h3511W,2008ApJ...672L..87E}, template fitting method~\citep{2012MNRAS.420.2162F}, the maximum entropy method~\citep{Gold_2009}, and Markov Chain Monte Carlo (MCMC) method~\citep{Gold_2009}, which rely on knowledge of foreground models, frequency dependence, and spectral energy distribution. In contrast, blind methods, like Independent Component Analysis (ICA)~\citep{2003MNRAS.344..544M,2002MNRAS.334...53M,2010MNRAS.402..207B}, Internal Linear Combination (ILC) ~\citep{10.1093/mnras/281.4.1297}, and Correlated Component Analysis (CCA)~\citep{2005EJASP2005.2400B}, extract cosmological signals solely from observed data, without assuming any specific foreground characteristics. Here, in this work, we extend the ILC method to 21 cm intensity mapping experiments. Specifically we employ harmonic space ILC  which is capable of addressing the challenges posed by 21 cm observations, allowing for more accurate extraction of cosmological signals from the radio sky.

The sky observations $d_\nu(p)$ at a frequency $\nu$ and pixel (or direction of the sky) $p$ can be written as a linear combination of the foregrounds, 21 cm signal  and instrumental noise,
\begin{equation}
    d_\nu(p) = s_\nu(p) + f_\nu(p) + n_\nu(p)\,,
    \label{model}
\end{equation}
where $f_\nu$ is the total foreground emissions at frequency $\nu$, $s_\nu$ is the instrumental noise in this channel and $s_\nu$ is the 21 cm signal at the same frequency channel that we aim to recover.  Each of these maps is assumed to have same HEALPix pixel resolution  $N_{\textrm {side}} = 128$. 

If there are $n_{ch}$ number of frequency channels, the net signal in the harmonic space, at a given frequency band $\nu$ can be written in thermodynamic temperature unit as,
\begin{eqnarray}
a^{\nu}_{\ell m} = a^{s(\nu)}_{\ell m} + a^{f(\nu)}_{\ell m} + a^{n(\nu)}_{\ell m}\,.
\label{freq_map}
\end{eqnarray}
The spherical harmonic coefficients $a^{\nu}_{\ell m}$ can be obtained from the pixel space signal $d_\nu(p)$ by performing a spherical harmonic transform:
\begin{equation}
    a^\nu_{\ell m} = \int d_\nu(p)Y_{lm}^{*}(p)d\Omega\, ,
    \label{aa1}
\end{equation}
where $Y_{lm}^{*}(p)$ are the complex conjugate of the spherical harmonic functions $Y_{lm}^{ }(p)$ and $d\Omega$ is the solid angle element in the direction $p$. Using Equation~\ref{model}, we can create $n_{ch} \times n_{ch}$ covariance matrix of the $n_{ch}$ input frequency maps for multipole $\ell$ as,
\begin{align}
\label{ocov}
\begin{aligned}
\widehat{\mathbf{R}}(\ell) &= \widehat{\mathbf{R}}_{\mathrm{21\,cm}}(\ell) + \widehat{\mathbf{R}}_n(\ell) + \widehat{\mathbf{R}}_{f}(\ell) \\
\end{aligned}
\end{align}
where $\widehat{\mathbf{R}}_{\mathrm{21\,cm}}$, $\widehat{\mathbf{R}}_n$  and $\widehat{\mathbf{R}}_{f}$  represents the $n_{ch} \times n_{ch}$  empirical covariance matrix of the 21 cm, noise and foreground respectively for the multipole $\ell$. The $(i, j)$th element of the empirical covariance matrix is given by,
\begin{eqnarray}
R^{ij}_{\ell} = \sum_{m=-l}^{\ell}\frac{a^i_{\ell m}a^{j*}_{\ell m}}{2\ell+1}\,.
\label{cross_cov}
\end{eqnarray}  

In a well-designed telescope system, the noise covariance matrix  $\widehat{\mathbf{R}}_n$ can be approximated as diagonal, implying that the noise is uncorrelated across frequency channels. This noise is difficult to distinguish from the 21 cm signal $\widehat{\mathbf{R}}_{\mathrm{21\,cm}}$, as both have sparse, diagonal-dominant structures. So, we combine $\widehat{\mathbf{R}}_n$ and $\widehat{\mathbf{R}}_{\mathrm{21 cm}}$ into an effective 21 cm signal covariance matrix $\widehat{\mathbf{R}}_{s}$, and focus on separating the foregrounds and the effective 21 cm signal. So here, we separate the foregrounds from the 21 cm signal and the instrumental noise. Using Equation~\ref{ocov}, we can write the effective 21 cm signal as,
\begin{equation}
\widehat{\mathbf{R}}_s(\ell) = \widehat{\mathbf{R}}_{\mathrm{21\,cm}}(\ell) + \widehat{\mathbf{R}}_n(\ell)\,.
 \label{eff_21 cm}
\end{equation}
Then using Equation~\ref{eff_21 cm}, we can rewrite  Equation~\ref{ocov} as,
\begin{equation}
\widehat{\mathbf{R}}(\ell) = \widehat{\mathbf{R}}_s(\ell) + \widehat{\mathbf{R}}_{f}(\ell)\,.
\label{cov_eff}
\end{equation}
\begin{figure}
\hspace*{-0.5cm}
	\includegraphics[scale=0.76,angle =00, trim={0cm 2cm 0cm 0cm}, clip]{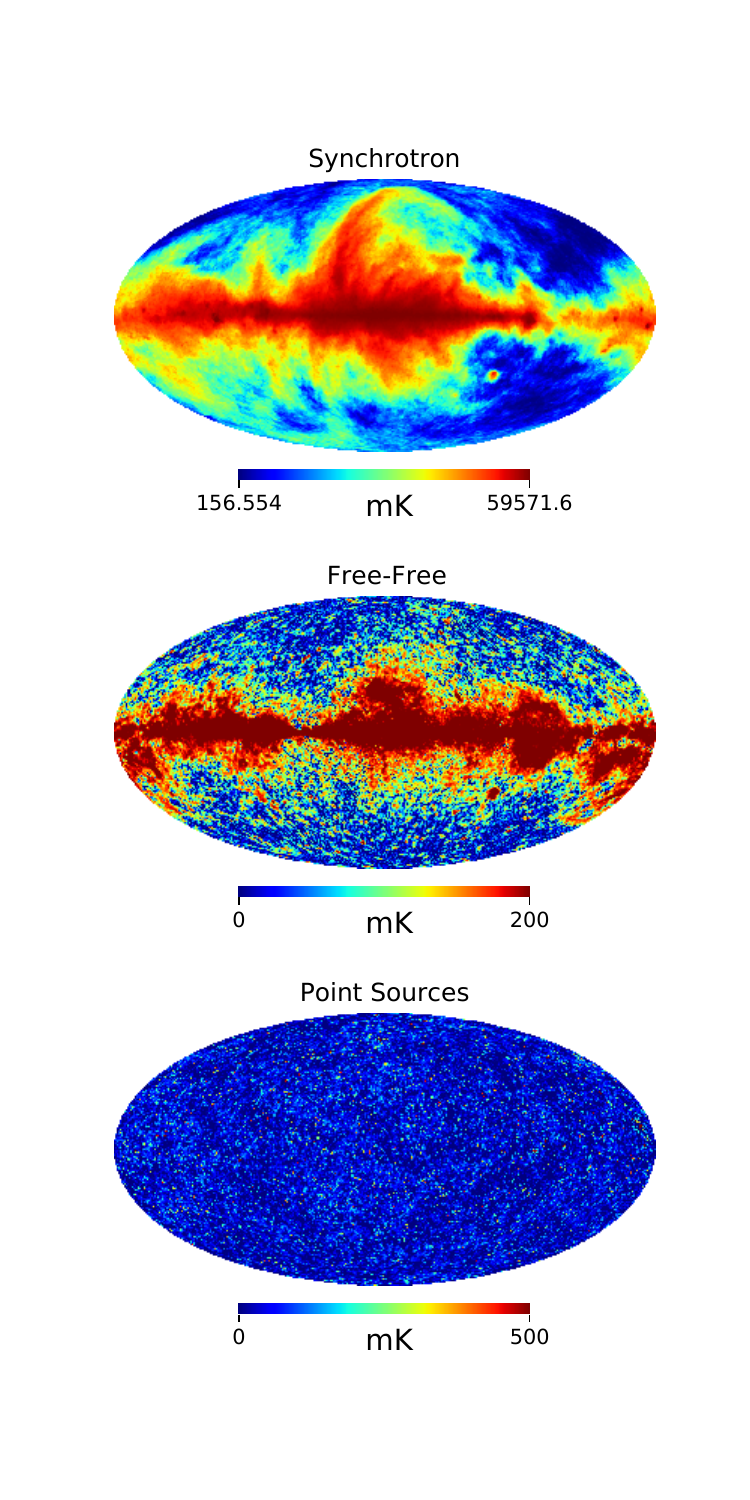}
	\caption{The top panel shows the synchrotron emission template at 1155 MHz corresponding to the {\tt s1} model of \texttt{PySM3}, and the middle panel represents the free-free emission template at 1155 MHz corresponding to the {\tt f1} model of \texttt{PySM3}. Here, both synchrotron and free-free emissions are plotted in histogram-equalized color scale. The bottom panel shows simulated extragalactic point sources at 1155 MHz from {\tt CORA}. We have limited the maximum pixel values of the free-free emission and point sources to 200 mK and 500 mK, respectively, although in the original maps these maximum pixel values extend to 128429 mK and 33546.7 mK. }
	\label{fig:templates}
\end{figure}
After obtaining the data covariance matrix, we apply a moving average to the data covariance matrix with a window of size $\pm 15$ in multipole space to account for local variations. 
This process effectively mitigates local variations in the data. 

The foregrounds in the observed sky may vary significantly with the observed directions in the sky, and their relative power ratio with the 21 cm signal  changes with angular scale. Theoretically, describing the foreground emissions would require a large number of degrees of freedom, potentially infinite for an infinitely narrow beam. However, in practice, we are limited by the number of frequency channels $n_{ch}$. The foreground components are correlated over frequencies and can be represented as a linear combination of $m$ independent templates, where $m$ is the dimension of the foreground subspace. Similarly, the 21 cm signal is partially correlated over adjacent frequencies (redshift bins) and can be described by $n_{ch} - m$ degrees of freedom. So we can write the 21 cm signal as a superposition of $n_{ch} - m$ independent templates $\mathbf{t}$,
\begin{equation}
\mathbf{s}_{\ell m} = \mathbf{S}(\ell) \mathbf{t}_{\ell m}, 
\label{basis}
\end{equation}
where $\mathbf{S}(\ell)$ is an $n_{ch} \times (n_{ch} - m)$ mixing matrix, and $\mathbf{t}$ represents $(n_{ch} - m)$ independent basis templates which allow us to explore the subspace dominated by the 21 cm signal and noise. The mixing matrix $\mathbf{S}(\ell)$ gives the contribution of each template to the 21 cm signal in each frequency channel.
Further using Equation~\ref{basis}, the empirical covariance matrix which incorporates the 21 cm signal and noise, can be written as an $n_{ch} \times n_{ch}$ matrix with rank $n_{ch} - m$:
\begin{equation}
\widehat{\mathbf{R}}_{s}(\ell) = \mathbf{S}(\ell)\widehat{\mathbf{R}}_t(\ell)\mathbf{S}^T(\ell),
\label{mix1}
\end{equation}
where $\widehat{\mathbf{R}}_t(\ell) = \langle\mathbf{t}\mathbf{t}^T\rangle$ is a full-rank $(n_{ch} - m) \times (n_{ch} - m)$ matrix.

Equation~\ref{mix1} shows that the mixing matrix $\mathbf{S}(\ell)$ maps the $n_{ch} - m$ dimensional subspace of the templates $\mathbf{t}$ to the full $n_{ch}$ dimensional space of the observed data. The mixing matrix $\mathbf{S}(\ell)$ relates the covariance structure of the templates in the reduced subspace to the covariance structure of the 21 cm signal in the full observation space. This effectively projects the $(n_{ch} - m) \times (n_{ch} - m)$ covariance matrix of the templates $\widehat{\mathbf{R}}_t$ onto the full $n_{ch} \times n_{ch}$ observation space.
\begin{figure}
	\centering
	\includegraphics[width=0.5\textwidth]{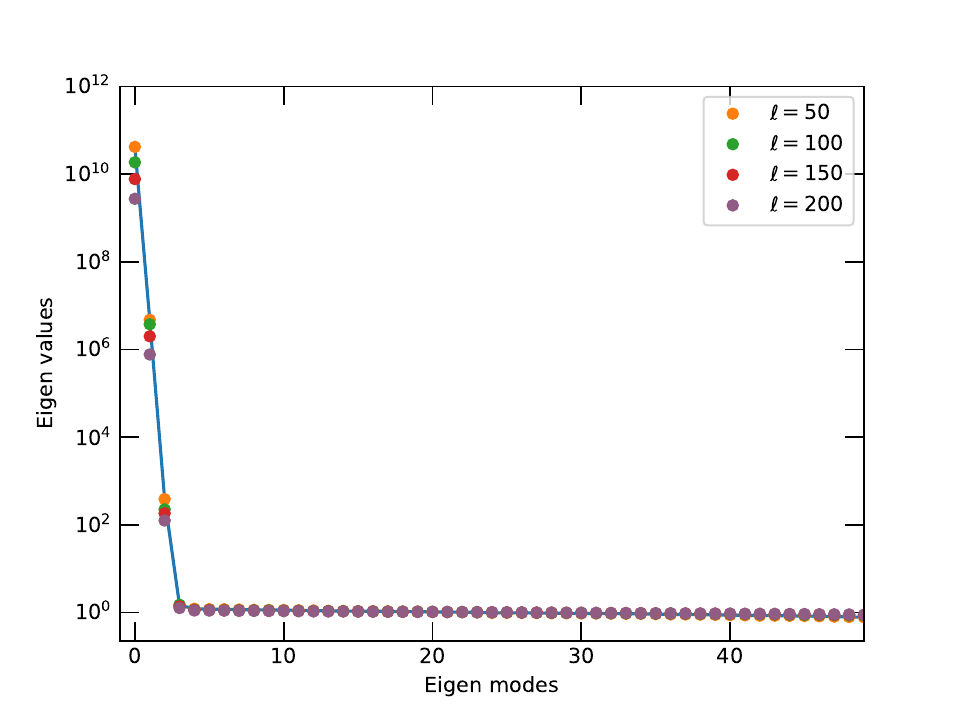}
	\caption{\small Eigenvalues of the normalized empirical covariance matrix given in Equation~\ref{emp_cov}. The three largest eigenvalues correspond to the modes dominated by foreground emissions, while the remaining eigenvalues close to unity represent the subspace spanned by the 21 cm signal and  noise.\normalsize
	}
	\label{fig:EV}
\end{figure}

\begin{figure*}
\hspace*{-2cm}
        \includegraphics[scale=0.43]{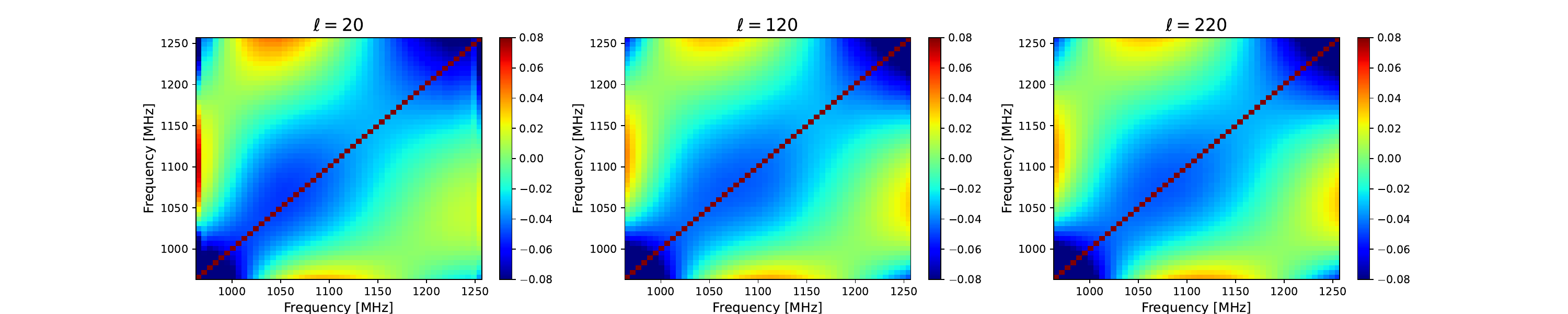}
        \caption{The harmonic space ILC weight matrices for different multipoles ($\ell$ = 20, 120, and 220) in the observed frequency range 960-1260 MHz. The structure of the weight matrices varies with the angular scale (multipole), indicating that the harmonic space ILC method adapts to the scale-dependent behavior of the foregrounds. 
        }
        \label{fig:weight_plot}
\end{figure*}

Now to estimate the mixing matrix, we first normalize the empirical covariance matrix by:
\begin{equation}
\label{emp_cov}
\mathbf{R}_{\mathrm{s}}^{-1/2}(\ell) \widehat{\mathbf{R}}(\ell)\mathbf{R}_{\mathrm{s}}^{-1/2}(\ell)\,,
\end{equation}
where $\mathbf{R}_{\mathrm{s}}$ is the theoretical 21 cm covariance matrix plus the noise covariance matrix, and $\widehat{\mathbf{R}}$ is the empirical covariance matrix obtained using Equation~\ref{cov_eff}. Using Equation~\ref{cov_eff}, we can decompose the normalized empirical covariance matrix as,
\begin{eqnarray}
\small
\nonumber
\mathbf{R}_{\mathrm{s}}^{-1/2}(\ell) \widehat{\mathbf{R}}(\ell)\mathbf{R}_{\mathrm{s}}^{-1/2}(\ell) 
&=& \mathbf{R}_{\mathrm{s}}^{-1/2}(\ell) \widehat{\mathbf{R}}_f(\ell)\mathbf{R}_{\mathrm{s}}^{-1/2}(\ell) \\
&& \hspace{-1.0cm}+\, \mathbf{R}_{\mathrm{s}}^{-1/2}(\ell) \widehat{\mathbf{R}}_s(\ell)\mathbf{R}_{\mathrm{s}}^{-1/2}(\ell)\,.
\label{tcov}
\end{eqnarray}

\begin{figure}
	\centering
	\includegraphics[width=0.5\textwidth]{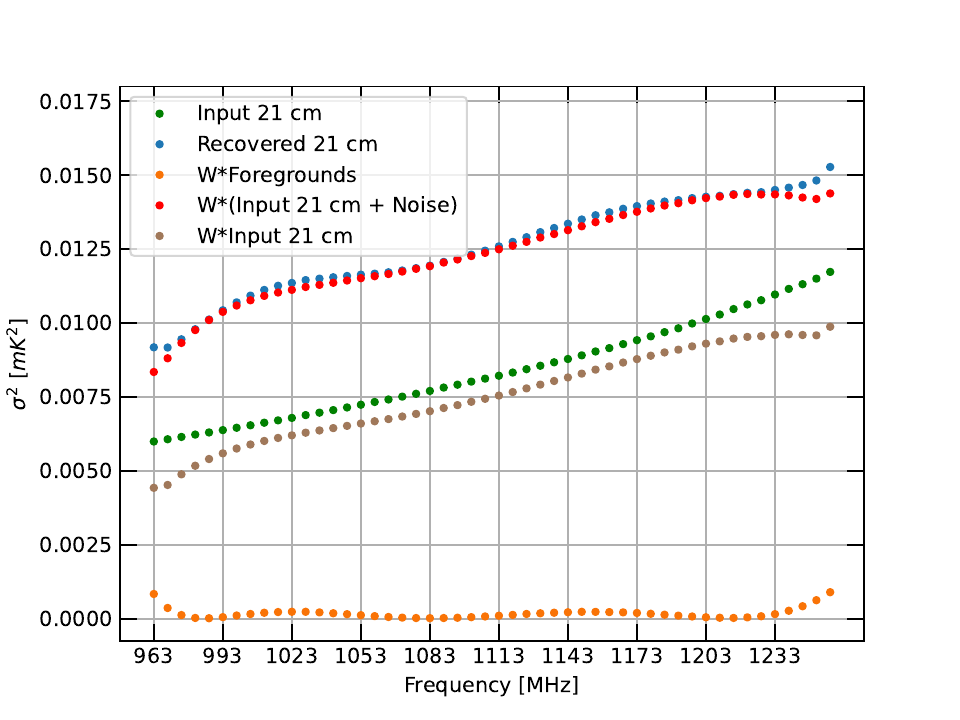}
	\caption{\small The figure demonstrates the effectiveness of the harmonic space ILC method in reconstructing the 21 cm signal across the observed frequency range. The vertical axis represents the mean variances obtained from 200 simulations. The strong suppression of the foreground residuals (orange) highlights the effectiveness of the method in separating the weak 21 cm signal from the dominant foreground contamination.
\normalsize
	}
	\label{fig:efficiency}
\end{figure}

Under the assumption that the prior theoretical 21 cm covariance matrix plus noise covariance matrix is close to the real 21 cm covariance matrix plus noise covariance matrix, we can rewrite Equation~\ref{tcov} as,
\begin{equation}
\label{power}
\mathbf{R}_{\mathrm{s}}^{-1/2}(\ell) \widehat{\mathbf{R}}(\ell)\mathbf{R}_{\mathrm{s}}^{-1/2}(\ell) = \mathbf{R}_{\mathrm{s}}^{-1/2}(\ell) \widehat{\mathbf{R}}_f(\ell)\mathbf{R}_{\mathrm{s}}^{-1/2}(\ell) + \tilde{\mathbf{I}}(\ell), 
\end{equation}
with $\tilde{\mathbf{I}}(\ell) \simeq \mathbf{I}$, where $\mathbf{I}$ is the identity matrix. By diagonalizing the normalized empirical covariance matrix we can isolate the subspace spanned by the 21 cm signal and noise. The eigenvectors corresponding to the eigenvalues close to one represent the basis templates that contribute to the 21 cm signal (+ noise). Using these basis templates and after some algebra, we can estimate the mixing matrix $\mathbf{S}$ as~\citep{2016MNRAS.456.2749O},
\begin{equation}
    \mathbf{S}(\ell) = \mathbf{R}_{\mathrm{s}}^{1/2}(\ell) \mathbf{U}_s(\ell)\,.
    \label{mix}
\end{equation}
where $\mathbf{U}_s$ is the eigenvectors representing the subspace spanned by the 21 cm signal (noise).

Now, the cleaned 21 cm signal can then be estimated by the linear combination of the input frequency channels,
\begin{equation}
\mathbf{s}_{\ell m} = \mathbf{W}(\ell)\mathbf{a}_{\ell m}\,.
\label{wt_1}
\end{equation}
where the ILC weight $\mathbf{W(\ell)}$ is an $n_{ch} \times n_{ch}$ matrix and $\mathbf{a}_{\ell m}$ is the vector containing the spherical harmonic modes of the observed map across the frequency channel for a given mode $\ell , m$. The matrix $\mathbf{W(\ell)}$ minimizes the total variance of the estimated vector $\mathbf{s}$, under the condition $\mathbf{W}(\ell)\mathbf{S}(\ell) = \mathbf{S}(\ell)$, can be obtained following Lagrange’s multiplier approach and given by ~\citep{2016MNRAS.456.2749O},
\begin{equation}
\mathbf{W}(\ell) = \mathbf{S}(\ell)[\mathbf{S}^T(\ell)\widehat{\mathbf{R}}^{-1}(\ell)\mathbf{S}(\ell)]^{-1}\mathbf{S}^T(\ell)\widehat{\mathbf{R}}^{-1}(\ell)\,.
\label{wgt}
\end{equation}
So following Equation~\ref{wt_1}, we can then define the cleaned 21 cm map in the harmonic space by forming a linear superposition of all input frequency maps:
\begin{eqnarray}
s^{\textrm {\textit{i}, Clean}}_{\ell m} = \sum_{j=1}^{n_{ch}} w^{ij}_{\ell}a^j_{\ell m}\,,
\label{CMAP}
\end{eqnarray}
where $w^{ij}_{\ell}$ represents the $(i,j)$th element of the weight matrix given in Equation \ref{wgt} and $s^{\textrm {\textit{i}, Clean}}_{\ell m}$ denotes the cleaned harmonic modes corresponding to a frequency channel $i$. The spherical harmonic coefficients $a^j_{\ell m}$ are obtained following Equation~\ref{aa1}.
Once the cleaned harmonic modes are obtained, the cleaned 21 cm map in pixel space can be reconstructed by performing an inverse spherical harmonic transform:
\begin{equation}
    \mathbf{s} ^i({p}) = \sum_{\ell > 0}\sum_{m=-\ell}^{\ell} s_{\ell m}^{\textrm {\textit{i}, Clean}} Y_{\ell m}({p}) \, .
    \label{cmap_1}
\end{equation}
Although these cleaned maps are free from foreground contamination, the galactic regions are strongly contaminated by synchrotron and possibly free-free emissions. So we further mask these cleaned 21 cm maps based on the procedure given in Section \ref{cleaned_maps}. Then the partial-sky cleaned 21 cm power spectrum corresponding to a frequency channel $i$ and a multipole $\ell$ is given by:
\begin{equation}
\tilde{C_{\ell}}^{\textrm {\textit{i},Clean}} = \frac{1}{2\ell + 1} \sum_{m=-\ell}^{\ell} \tilde{s}^{\textrm {\textit{i}, Clean}}_{\ell m} \tilde{s}^{\textrm {\textit{i}, Clean}\ast}_{\ell m} \, .
\label{cmap_ps}
\end{equation}
Here, $\tilde{s}^{\textrm {\textit{i}, Clean}}_{\ell m}$ denote the pseudo-harmonics for a given frequency channel $i$, which can be obtained from the aforementioned cleaned and masked 21 cm maps by performing a spherical harmonic transform as in Equation~\ref{aa1}.

\begin{figure*}
\hspace*{-3cm}
    \includegraphics[scale=0.6, trim={-4.5cm 0cm 0cm 0cm}, clip]{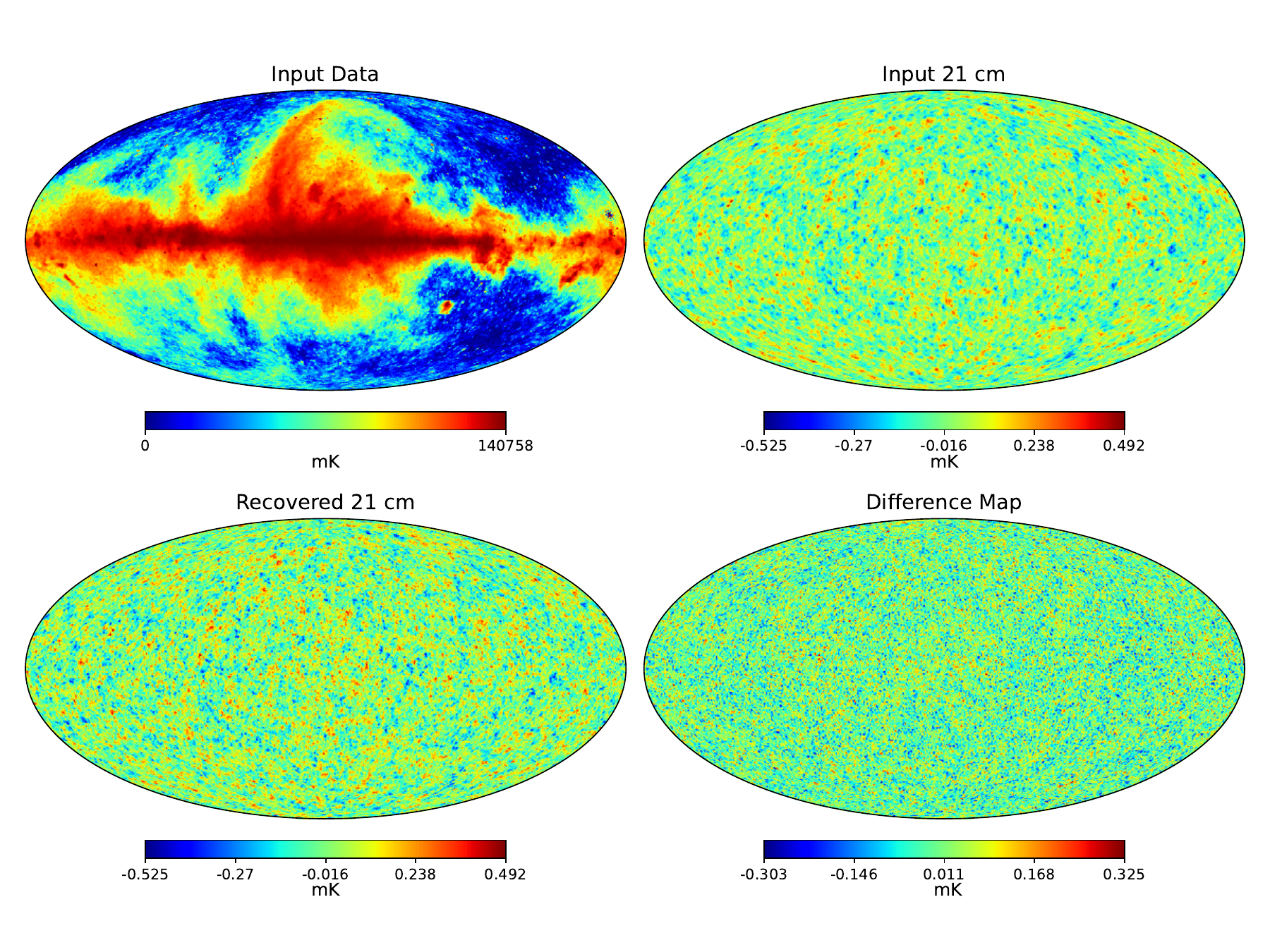}
        \caption{In the top left panel, we show the input data map comprising all foregrounds, 21 cm signal, and noise corresponding to the central frequency channel at 1209.0 MHz. The pure 21 cm signal map for this frequency channel is shown in the top right panel, and the recovered 21 cm signal map is shown in the bottom left panel. The bottom right panel depicts the residual contamination map obtained by subtracting the input pure 21 cm map from the corresponding recovered map. Here, the input data maps are plotted in the histogram equalized color scale. 
        } 
        \label{fig:maps}
\end{figure*}

As discussed earlier, after separating the foregrounds from the effective 21 cm signal (21 cm + noise), some specific noise debiasing is required from the cleaned partial-sky effective 21 cm signal to obtain the cleaned partial-sky 21 cm signal power spectrum. The debiased cleaned partial-sky 21 cm power spectrum can be obtained by:
\begin{equation}
    \tilde{C_{\ell}}^{\textrm {\textit{i},Clean}^{\prime}} = \tilde{C_{\ell}}^{\textrm {\textit{i},Clean}} - \sum_{j=1}^{n_{ch}} w^{ij}_{\ell} \tilde{\sigma_{\ell}}^{\textrm {\textit{jj}}}w^{ji}_{\ell}\,,
\end{equation}
where $\tilde{\sigma_{\ell}}^{\textrm {\textit{jj}}}$ represents the partial-sky noise power spectrum for a given frequency channel $j$ which are estimated by generating Monte Carlo realizations of the noise maps after applying the masks as discussed in Section \ref{cleaned_maps}. Moreover, $w^{ij}$ denote the $(i, j)$th elements of the weight matrix given in Equation \ref{wgt}. After the noise bias subtraction, our debiased partial-sky 21 cm power spectrum is given by $\tilde{C_{\ell}}^{\textrm {\textit{i}, Clean}^{\prime}}$. However, for the purpose of common usage of notations, we represent debiased partial-sky 21 cm power spectrum by $\tilde{C_{\ell}}^{\textrm {\textit{i}, Clean}}$.

\begin{figure*}
\hspace*{-3cm}
        \includegraphics[scale=0.7, trim={0cm 1.5cm 0cm 0cm}, clip]{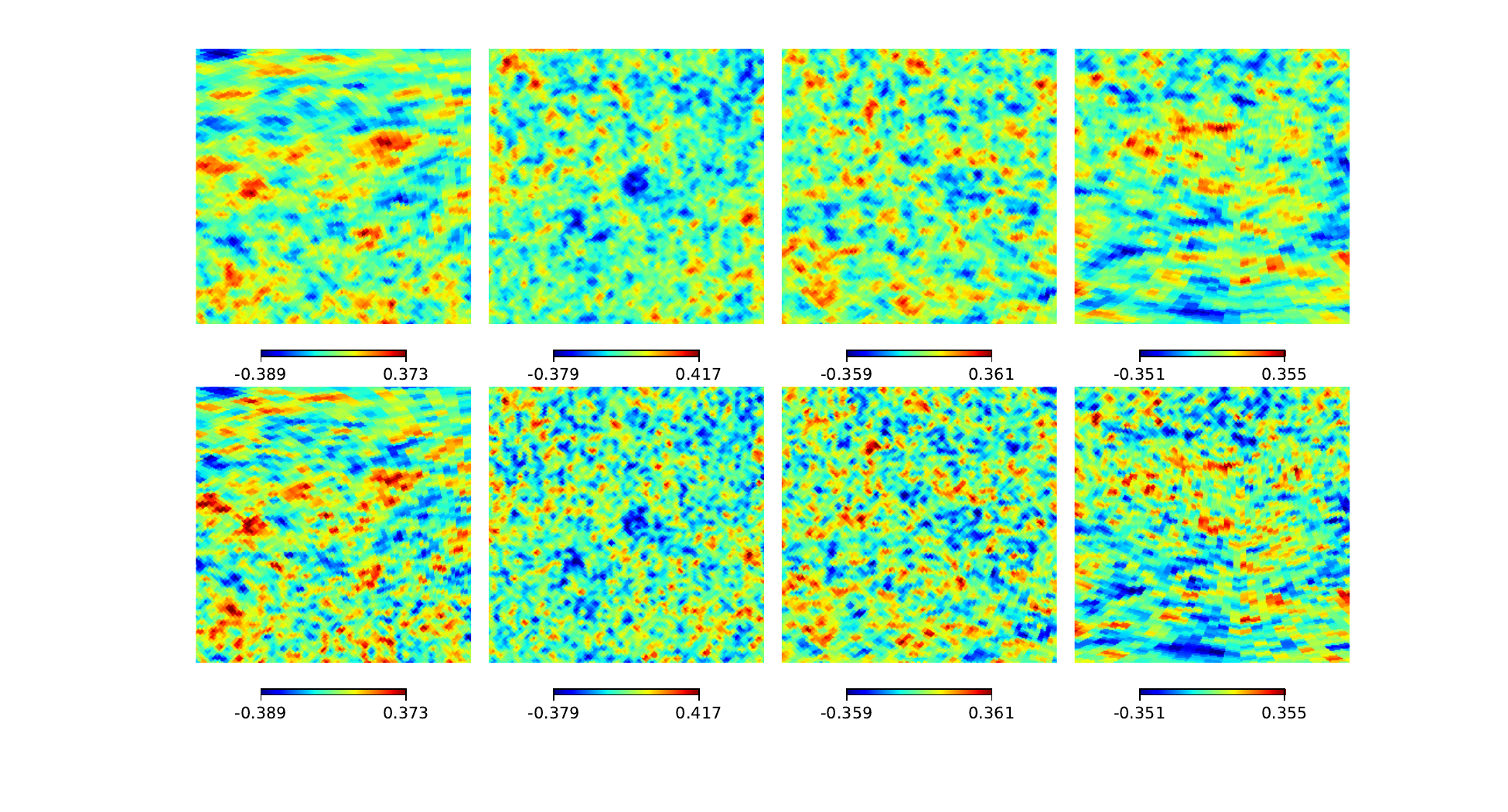}
        \caption{In the top panel, we show the input 21 cm map covering a $50^{\circ} \times 50^{\circ}$ patch of the sky at various sky positions for the 1209 MHz channel. The first column represents a patch of the sky near the North Pole, the second and third columns represent regions near the Galactic plane, and the last column represents the patch near the South Pole. The bottom panel shows the recovered 21 cm maps corresponding to these regions of the sky. There is a noticeable similarity between the input and recovered 21 cm maps in these regions.}
        \label{fig:cartview}
\end{figure*}
\begin{figure*}
	\vspace*{-1cm}  
	\hspace*{0cm}   
	\includegraphics[scale=0.045, trim={1.0cm 0cm 0cm 0cm}, clip]{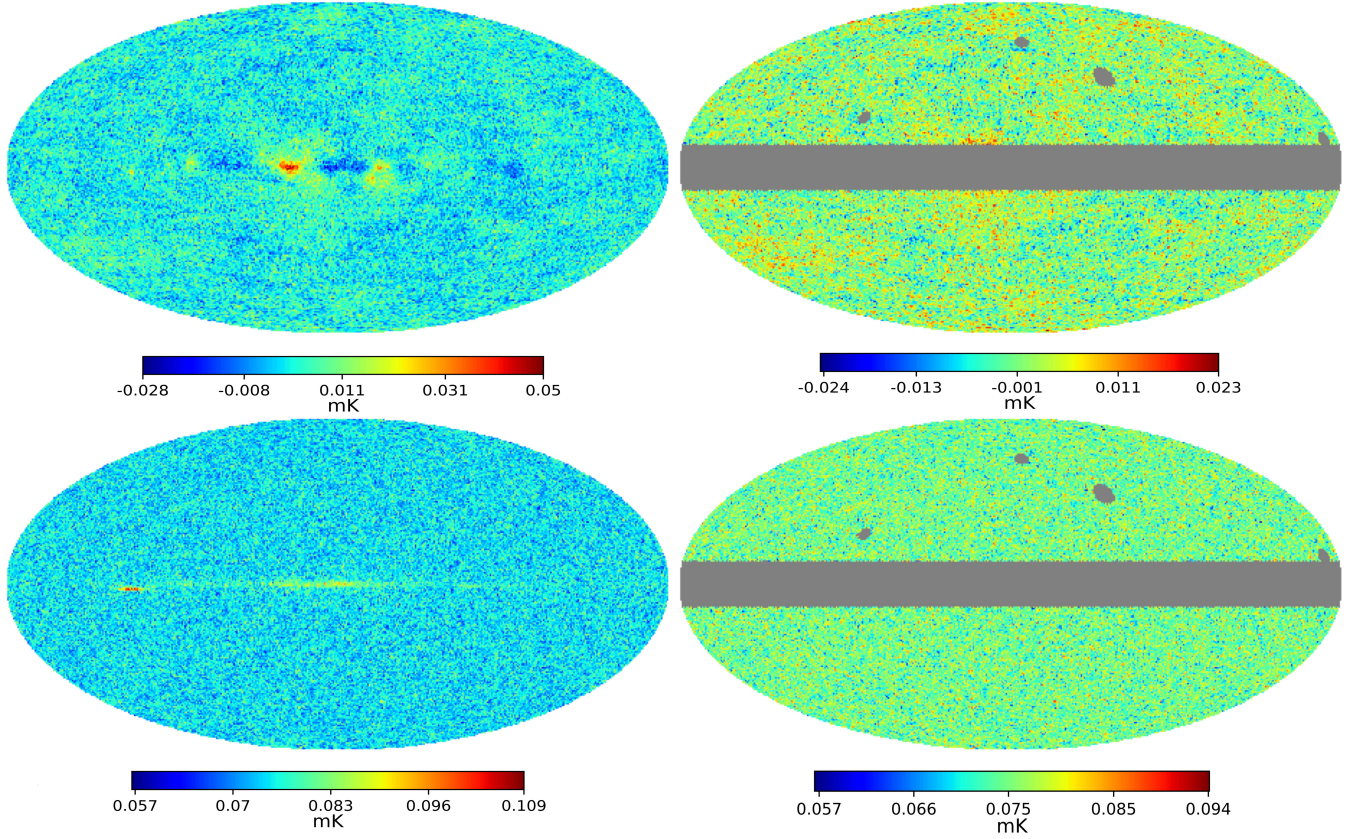}
	\caption{The top left panel shows the mean difference map obtained by subtracting the recovered 21 cm signal from the input 21 cm signal map over 200 simulations at a random frequency channel 1089 MHz. The bottom left panel depicts the standard deviation map obtained from 200 difference maps. The residuals are mostly present near the Galactic plane due to strong foreground contamination in that region. The top right and bottom right panels show the mean difference and standard deviation maps after masking the Galactic plane and a few bright point sources above the Galactic plane.}
	\label{fig:sdmaps_full}
\end{figure*}

Once we estimate the partial-sky cleaned 21 cm angular power $\tilde{C_{\ell}}^{\textrm {\textit{i}, Clean}}$ for a given frequency channel $i$, we estimate the full-sky 21 cm angular  power spectrum by implementing the MASTER algorithm~\citep{2002ApJ...567....2H} (see the Section~\ref{coupling_matrix} for more details), such that:
\begin{equation}
{C_{\ell}}^{\textrm {\textit{i},Clean}} = M_{\ell\ell^\prime}^{-1,i}\tilde{C_{\ell^{\prime}}}^{\textrm {\textit{i},Clean}}  \, .
\label{reccl_1}
\end{equation}
Here, the mode-mode coupling matrix $M_{\ell\ell^\prime}$ is given by:
\begin{equation}
M_{\ell\ell^\prime} = \frac{2\ell^\prime + 1}{4\pi} \sum_{\ell^{\prime\prime}} (2\ell^{\prime\prime} + 1) C_{\ell^{\prime\prime}}^{M} \begin{pmatrix}
\ell & \ell^\prime & \ell^{\prime\prime} \\ 0 & 0 & 0
\end{pmatrix}^{2}\, ,
\label{mllp_1}
\end{equation}
where $C_\ell^M$ is the power spectrum of the Gaussian smoothed mask used in our analysis, and $\begin{pmatrix}
\ell & \ell^{\prime} & \ell^{\prime\prime} \\ 0 & 0 & 0
\end{pmatrix}$ is the Wigner 3j symbol. The Wigner 3j symbol are related  to the Clebsch-Gordan coefficients and describe the coupling of angular momenta.

\begin{figure*}
	\hspace*{-1.5cm}
		\includegraphics[scale=0.55]{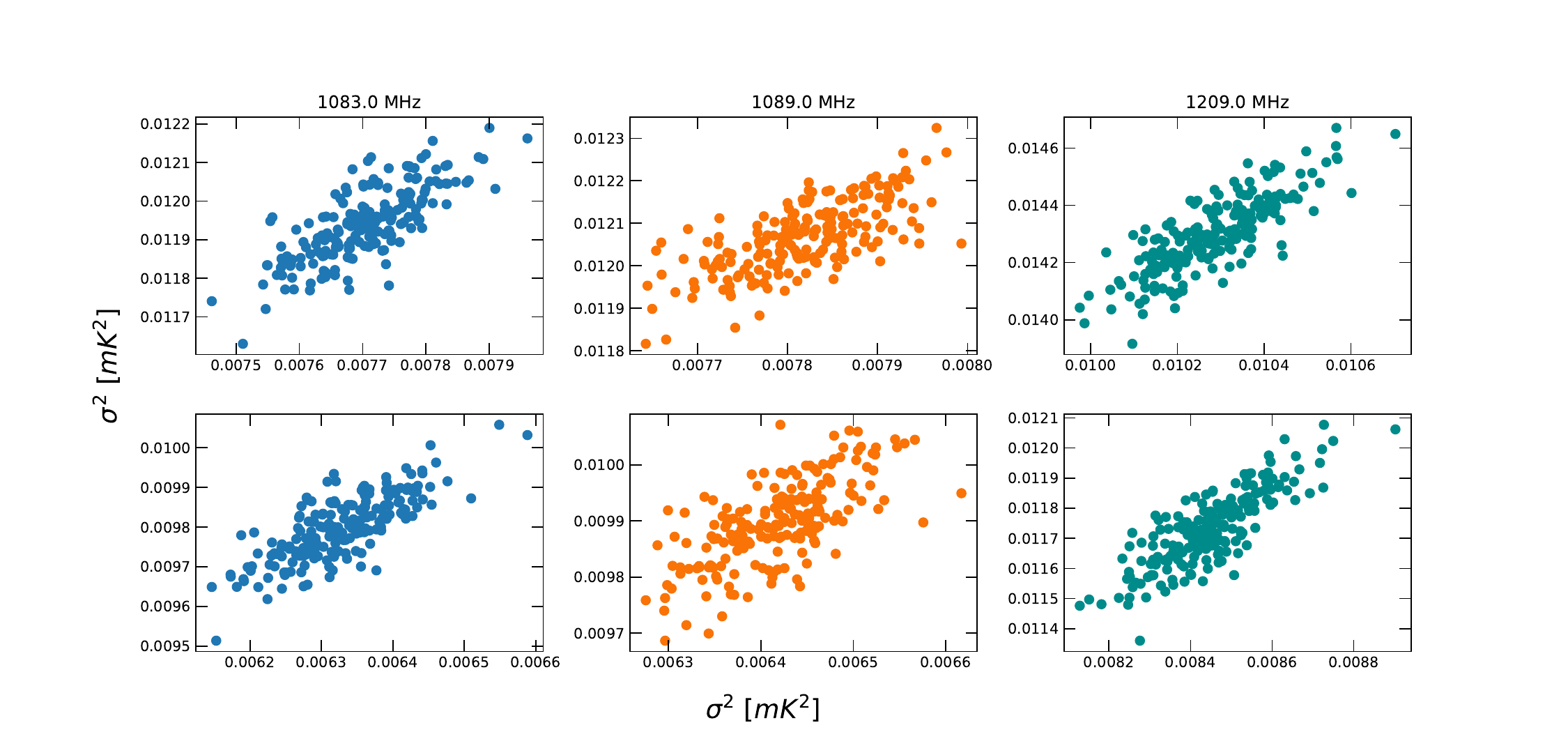}
		\caption{In the top panel, we show the scatter plots of the variances of the input pure 21 cm full-sky maps versus the corresponding recovered 21 cm signals for 200 simulations at various observed frequency channels. The bottom panel depicts the scatter plots of the variances after applying the mask to both the input and recovered maps. An interesting point to note is that the variance values over the partial sky are smaller than the full sky estimates, as expected.
		}
		\label{fig:var_scatter}
\end{figure*}


\section{Simulations}\label{sim}

In this work, we simulate the fixed foreground templates and randomly generate thermal noise in the frequency range of $960$ to $1250$ MHz, corresponding to the BINGO experiment~\citep{2013MNRAS.434.1239B}. This bandwidth corresponds to a redshift range of 0.13 to 0.48. The foreground and noise maps are then added to the respective HI signal to produce realistic sky emissions. In the following section, we discuss the various input templates that we use for the simulations.

\subsection{Foreground Model}
\label{fg}

The major astrophysical sources of emissions that can pose a significant challenge for the estimation of the HI signal and its interpretation are free-free, synchrotron, and point sources. We utilize the publicly available software package \texttt{PySM3}~\citep{2017MNRAS.469.2821T} to simulate the free-free and synchrotron emissions, while the templates for point sources are obtained using \texttt{CORA}~\citep{2014ApJ...781...57S}. The Figure~\ref{fig:templates} shows the major sources of astrophysical contamination which corrupt the detection of the 21 cm signal. A more detailed description of the individual foreground templates is discussed below.

\subsubsection{Synchrotron \texttt{s1}}

We utilize the \texttt{s1} template from \texttt{PySM3} to generate realistic synchrotron emission within the BINGO bandwidth. The synchrotron templates are derived from the 408 MHz Haslam map following~\cite{2015MNRAS.451.4311R}. The \texttt{s1} model in \texttt{PySM3} considers a power-law model with a spatially varying spectral index $\beta(\hat{n})$ at the WMAP reference frequency $\nu_{0} = 23$ GHz, given by,
\begin{equation}
    I^{synch}_{\nu}(\hat{n}) = A_{\nu_{0}}(\hat{n})\left( \frac{\nu}{\nu_{0}} \right)^{\beta(\hat{n})}\,.
\end{equation}
The spectral index map \texttt{s1} in \texttt{PySM3} is adopted from `Model 4' of \cite{2008A&A...490.1093M}.

\subsubsection{Free-free \texttt{f1}}

The free-free emission template in \texttt{PySM3} follows a power-law model with a constant spectral index of -2.14, which flattens abruptly at lower frequencies. This template employs degree-scale smoothed emission measure and effective electron temperature \texttt{Commander} templates~\citep{2016A&A...594A..10P}. The choice of spectral index aligns with WMAP and Planck measurements for electrons at $\sim$ 8000K~\citep{2016A&A...594A..10P}.

\subsubsection{Point Sources}

We employ \texttt{CORA}~\citep{2014ApJ...781...57S} to simulate extragalactic point sources within the BINGO bandwidth. The simulations of extragalactic point sources in \texttt{CORA} consist primarily of three components: a population of bright point sources (S $>$ 10 Jy at 151 MHz),  a synthetic population of fainter sources extending down to 0.1 Jy at 151 MHz, and an unresolved background of dimmer sources (S $<$ 0.1 Jy) modeled as a Gaussian random realization~\citep{2002ApJ...564..576D}.

\subsection{21 cm Emission}
For the simulations of the 21 cm signal, we use \texttt{CORA} package to generate the 21 cm signal in the BINGO bandwidth. In context of \texttt{CORA}, the simulations of the 21 cm signals are obtained by drawing Gaussian realizations from the flat-sky angular power spectrum~\citep{2014ApJ...781...57S}. Moreover for the HI signal, we assume the cosmological model given in Planck Collaboration~\citep{2014A&A...571A..16P}. On large scales, the 21 cm brightness temperature is a biased tracer of the matter density field~\citep{1987MNRAS.227....1K}. The power spectrum of the 21 cm emission is given by~\citep{1987MNRAS.227....1K},
\begin{equation} \label{eq:PTb}
  P_{T_{b}}(\vec{k}; z, z') = \bar{T}_{b}(z) \bar{T}_{b}(z') (b + f
  \mu^{2})^{2} P_{m}(k; z, z'),
\end{equation}
where $b$ is the bias, $f$ is the growth rate, and $P_{m}\left(k ; z, z^{\prime}\right)=P(k) D_{+}(z) D_{+}\left(z^{\prime}\right)$ is the real-space matter power spectrum, $D_{+}$ is the growth factor normalized such that $D_{+}(0)=1$. We consider the bias factor $b=1$ at all redshifts. The mean brightness
temperature can be represented as~\citep{2008PhRvL.100i1303C},
\small
\begin{equation}\label{eq:Tbz}
  \bar{T}_{b}(z) = 0.3 \, \left( \frac{\Omega_{\text{H{~\sc i}}}}{
      10^{-3}} \right)   \left( \frac{1 +  z}{2.5} \right)^{1/2}    
      \left[ \frac{\Omega_{\rm m} + (1 + z)^{-3}
      \Omega_{\Lambda}}{0.29} \right]^{-1/2}\,\text{mK}.
\end{equation}
\normalsize
The 21 cm angular power spectrum $C_\ell$ characterizes the fluctuations in 21 cm brightness temperature as a function of angular scale. However, the precise computation of this spectrum involves intricate double-integration over highly oscillatory functions. The 21 cm angular power spectrum is given by~\citep{2007MNRAS.378..119D},
\begin{equation}
    C_\ell(\Delta \nu) \propto \int k^2 \,dk \,j_\ell(k \chi) j_\ell(k \chi') P_{T_b}(k; z, z')\,,
\end{equation}
where $\Delta \nu = \nu' - \nu$  and  $\chi (\chi')$ represents the comoving distance corresponding to redshift $z(z')$, which is associated with the frequency $\nu (\nu')$. The \texttt{CORA} employs flat-sky approximation, where the 21 cm angular power spectrum can be approximated as,
\begin{equation}
    C_\ell(z, z') = \frac{1}{\pi \chi \chi'}\int_{0}^{\infty} dk_\parallel \cos(k_\parallel \Delta\chi) P_{T_b}(k; z, z')
\end{equation}
where $\Delta \chi = \chi -\chi'$. The vector $\textbf{k}$ is characterized by components $k_\parallel$ and $\ell/\bar \chi$ in the directions parallel and perpendicular to the line of sight, respectively (where $\bar \chi$ is the mean of $\chi$ and $\chi'$). This approximation remains accurate at the level of 1\% for $\ell > 10$, as demonstrated in~\citep{2007MNRAS.378..119D}.

\subsection{Thermal Noise}\label{noise_sec}

Inorder to simulate the instrumental noise, we consider the only thermal noise, assuming it adheres to a uniform Gaussian distribution across the celestial sphere. So in our analysis, we do not consider pink noise and other systematic error sources. Additionally, for simplicity, we assume a constant noise amplitude across frequency channels.
The sensitivity per pixel $\sigma$ for a single-dish telescope can be written as~\citep{2013tra..book.....W},
\begin{equation}
    \sigma = \frac{T_{sys}}{\sqrt{t_{pix}\Delta \nu}}\,,
\end{equation}
where $T_{sys}$ is the system temperature,  $\Delta \nu$ is the frequency channel width, and$t_{pix}$ is the integration time per pixel. The $t_{pix}$ depends on the number of feed horns $n_f$, total integration time $t_{obs}$, survey area $\Omega_{sur}$, and beam area $\Omega_{pix}$ through:
\begin{equation}
    t_{pix} = n_f t_{obs} \frac{\Omega_{pix}}{\Omega_{sur}}\,,
    \label{th_noise}
\end{equation}
with $\Omega_{pix} = \theta ^2_{\text{FWHM}}$, where $\theta_{\text{FWHM}}$ is the full width at half minimum of the beam in radian. We use the instrumental parameters compactible with those of the BINGO experiment, with the exception of the larger sky coverage, leading to an increased noise amplitude per pixel. Key parameters of our experiment are detailed in Table~\ref{tab:instrument}. 

\begin{table}
	\addtolength{\tabcolsep}{8pt}
	\renewcommand{\arraystretch}{2.0}
	\centering
	\begin{tabular} { l  c}
		\hline
		\hline
		Parameters &  \\
		\hline
		\hline
		{Redshift range [$z_{min} , z_{max}$ ]} & [0.13, 0.48]     \\
		{Bandwidth [$\nu_{min} , \nu_{max}$ ](MHz)} & [960, 1260]     \\
		{Number of feed horns $n_f$} & 80    \\
            Sky coverage $\Omega_{\text{sur}}$ ($\text{sr}^2$) & $4\pi$ \\
            Observation time $t_{\text{obs}}$ (yrs) & 1 \\
            {System temperature $T_{sys}$ (K)} & 50 \\
            {Beamwidth $\theta_{\text{FWHM}}$ (arcmin)} & $40^\prime$ (0.12 rad)\\
		\hline
		\hline
	\end{tabular}
	\caption{ Instrumental parameters for a single-dish simulation. }
	\label{tab:instrument}
\end{table}



\section{Methodology and Results}\label{results}
This section outlines the methodology and presents the results obtained after applying our method to the simulated foreground and noise-contaminated 21 cm signal map in the BINGO frequency bandwidth. 

\subsection{Cleaned Maps}\label{cleaned_maps}


We construct $50$ full-sky frequency maps of 21 cm as will be observed by BINGO
in the frequency range 960 to 1260 MHz. The spectral resolution ($\Delta \nu$) of each data
set is therefore 6 MHz. We add the pure 21 cm signal, synchtrotron, free-free, point source
and thermal noise maps (discussed in Section~\ref{sim}) to construct realistic frequency maps
of BINGO at HEALPix~\citep{2005ApJ...622..759G} pixel resolution parameter $N_{side} = 128$. We smooth each foreground and 21 cm
signal by a circularly symmetric Gaussian beam window function of FWHM = $40$ arcmin. The noise
maps are not smoothed before addition since the noise model (Equation~\ref{th_noise}) already incorporates a pixel uncorrelated
white noise model with noise standard deviation at each pixel reduced by the area of this Gaussian
window function.

To perform the foreground removal, we first convert each of the frequency maps to harmonic space coefficients up to
$\ell_{max} = 256$. Assuming the effect of beam smoothing is implicit (in the foreground and 21 cm components)
an observed map at the frequency $\nu$ in the harmonic space is given by a set of $a_{\ell m}$ coefficients as described by Equation~\ref{freq_map}. The weights for the linear combination of input frequency maps (Equation~\ref{CMAP}) are obtained using the mixing matrix $\bf S(\ell)$ and the empirical covariance matrix $\widehat{\mathbf{R}}(\ell)$ which can be obtained by using Equation~\ref{cross_cov}. The mixing matrix ${\bf S}(\ell)$ is determined using Equation~\ref{mix} and is related to the eigenvectors corresponding to the eigenvalues close to unity. This unit response ensures that we recover the proper subspace of the 21 cm signal and (relatively subdominant) instrumental noise. We perform 200 Monte Carlo simulations of the entire foreground cleaning procedure to assess the performance of the method statistically. To generate the random 21 cm signal realizations at all the 50 frequencies chosen in this work we use the theoretical 21 cm with the cosmological parameters consistent with the Planck~\citep{2020A&A...641A...6P}. The noise realizations were generated using the theoretical white noise model as described in Section~\ref{noise_sec}.

The Figure~\ref{fig:EV} shows the eigenvalues of the normalized empirical covariance matrix, where there is a significant difference between the eigenvalues spanned by the foregrounds and the 21 cm signal. The three largest eigenvalues correspond to the modes dominated by foreground emissions, while the remaining eigenvalues close to unity represent the subspace spanned by the 21 cm signal and noise. Furthermore, in Figure~\ref{fig:weight_plot}, we show weight matrices for different multipoles ($\ell$ = 20, 120, and 220) in the observed frequency range 960-1260 MHz. The varying structure of the weight matrices with angular scale demonstrates that the harmonic space ILC method adapts to the scale-dependent behaviour of the foregrounds. Once the cleaned map is obtained in harmonic space at a frequency $\nu_i$, the pixel space map is obtained by a forward spherical harmonic transformation (Equation~\ref{cmap_1}).

\begin{figure*}
        \includegraphics[scale=0.55]{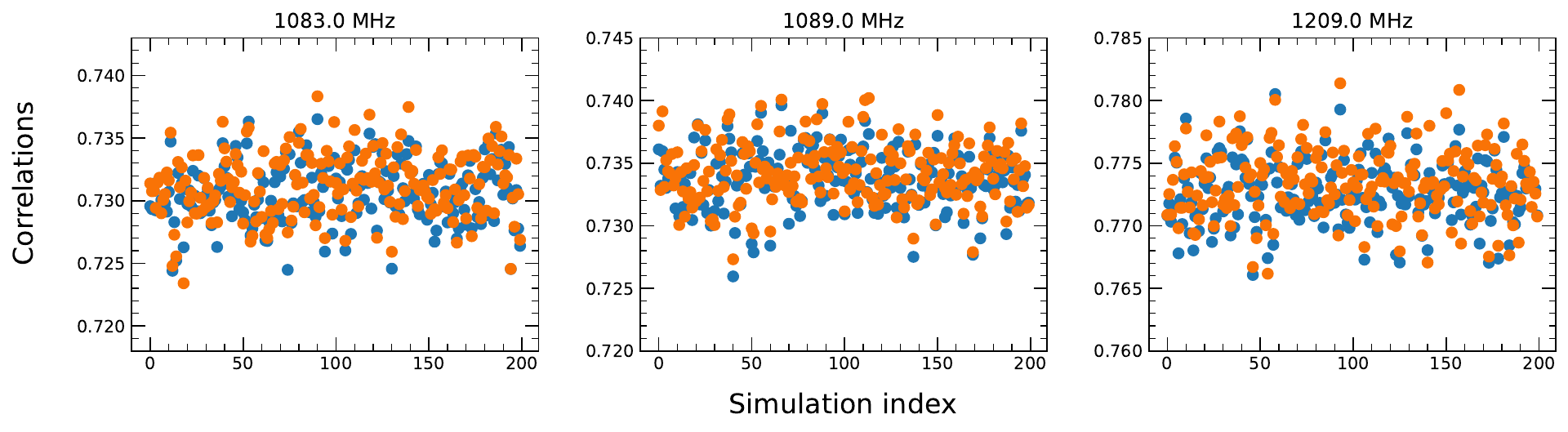}
        \caption{We show the correlations between the input pure 21 cm maps and the corresponding recovered 21 cm signals for 200 simulations at various observed frequency channels. The blue points represent the correlations for the full-sky scenario, while the orange points represent the correlations for the masked scenario.   Overall, the correlations  between the cleaned 21 cm signal and corresponding input 21 cm frequency map takes similar value for both full sky and partial sky.  This is consistent with the assumption of statistical isotropy of the 21 cm signal (and detector noise) used in our work.
        }
        \label{fig:corr_scatter}
\end{figure*}
\begin{figure}
	\hspace*{-0.5cm}
		\includegraphics[scale=0.48]{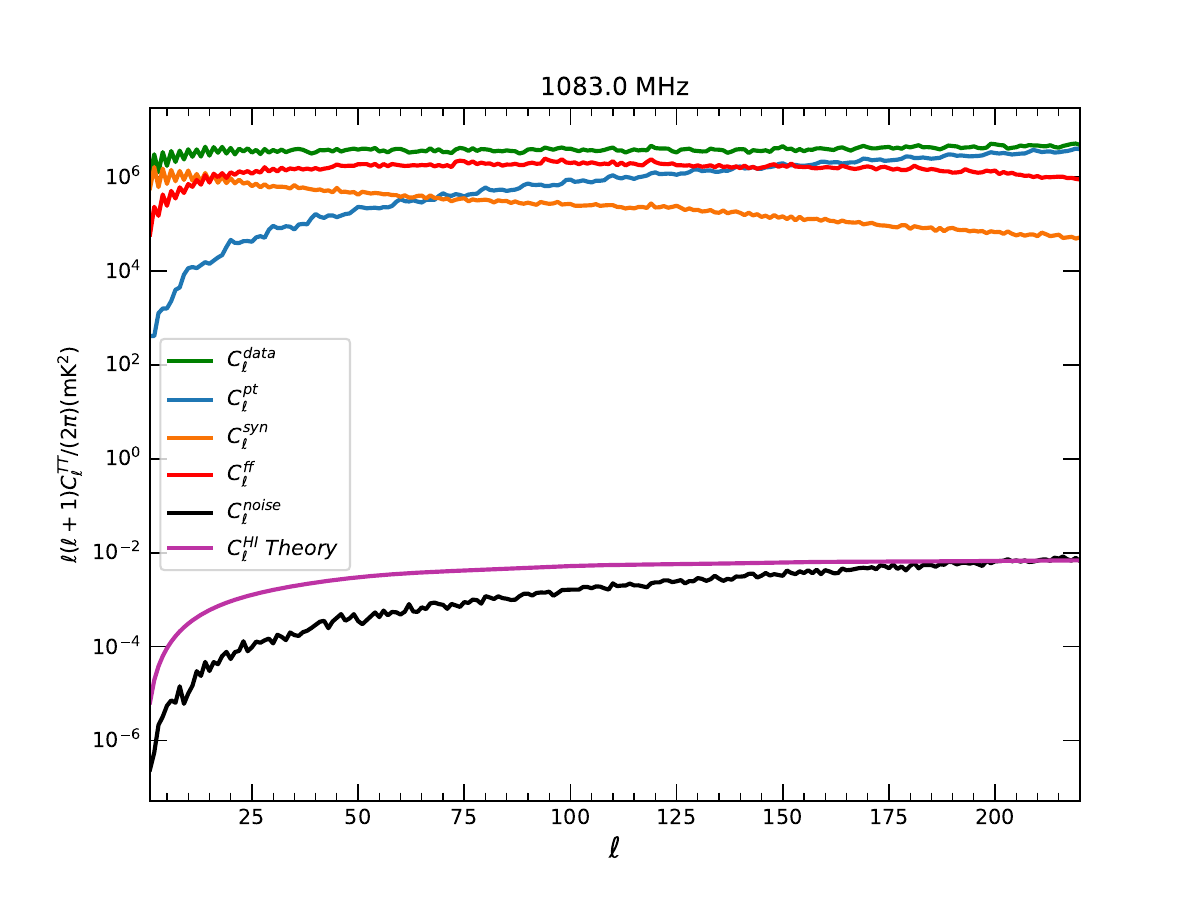}
		\caption{The power spectra of various astrophysical sources of contamination and thermal noise in comparison with the theoretical 21 cm power spectrum for the central frequency channel 1083.0 MHz. }
		\label{fig:cl_all}
\end{figure}
In Figure~\ref{fig:efficiency} we show the effectiveness of the harmonic space ILC in combination with PCA in reconstructing the 21 cm signal across the observed frequency range. We compute the variance of pure 21 cm maps at each frequency and for all $200$ Monte Carlo simulations. For each frequency, we then form average variance using $200$ Monte Carlo simulations. The mean variance of input pure 21 cm maps obtained from 200 Monte Carlo simulations is shown in green, while the mean variances computed similarly from the recovered 21 cm signal are shown in blue. As we remove only foregrounds from the mixture of foregrounds, 21 cm signal and noise, the variance of the recovered 21 cm signal is higher than the input pure 21 cm signal, as expected, since the recovered maps contain noise. We can, however, compute the mean variance due to the 21 cm component in the cleaned maps. We do so by linearly combining pure 21 cm maps at each frequency by  the corresponding ILC weights  in the harmonic space at each $\ell$. We then compute mean variance of the resulting maps from $200$ Monte Carlo simulations at each frequency. We show these variances as the gray points in Figure~\ref{fig:efficiency}. The grey curve remains somewhat below the green since the PCA method retains only a subset of eigenvalues. In a similar fashion, the red curve represents variance when pure 21 cm and noise maps are linearly combined with the ILC weights. The slight deviation between the green and red variances can be attributed to minor residual foregrounds left in the recovered maps. To quantify the foreground residuals, we multiply the ILC weights with the input foreground templates and obtain the mean variances from 200 simulations. The strong suppression of the foreground residuals (orange) highlights the effectiveness of the method in separating the weak 21 cm signal from the dominant foreground contamination. It is interesting to note that the residual foreground shows the presence of clear and minimum variances at the frequency channels 1083.0, 1089.0,  1209.0 MHz and neighbouring channels. Based on the higher effectiveness of the signal recovery, we showcase the performance of our methodology at these $3$  frequency channels in the forthcoming discussions. The results of the cleaned maps are shown in Figure~\ref{fig:maps}.

\begin{figure*}[!htbp]
	\subfloat{\includegraphics[width=.5\linewidth]{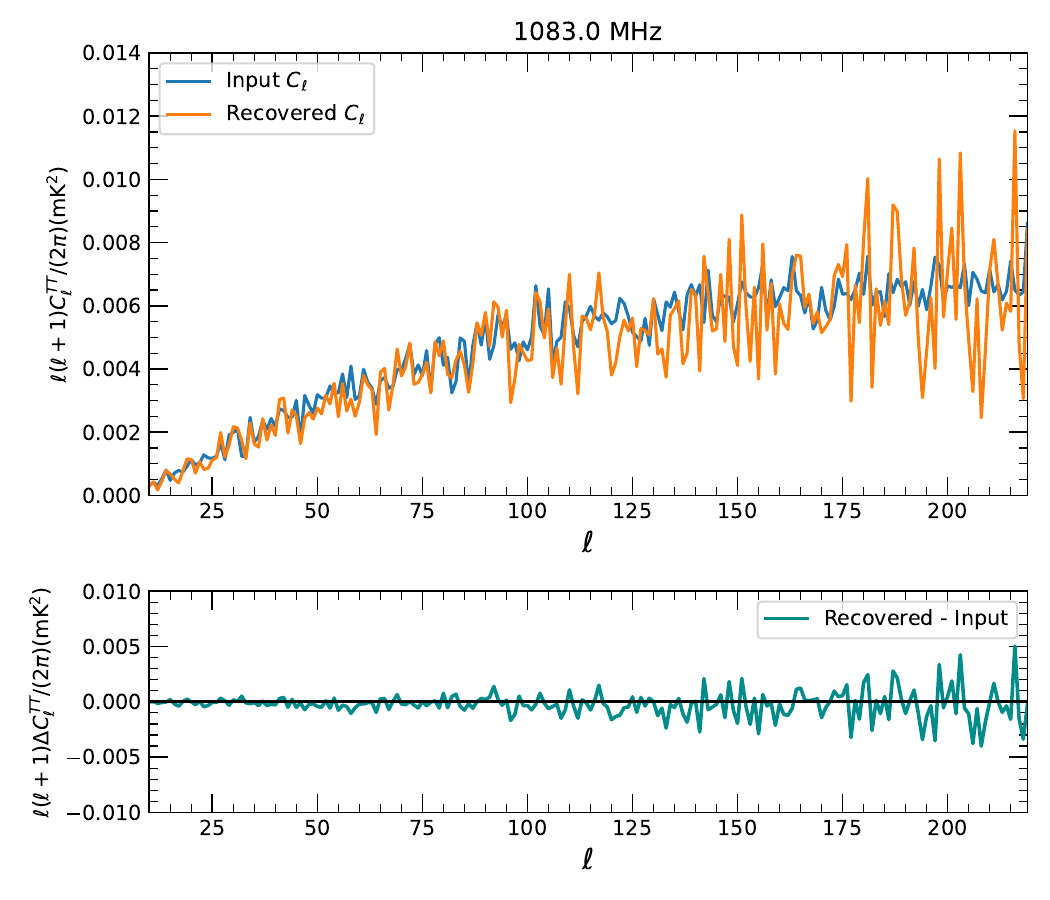}}\hfill
	\subfloat{\includegraphics[width=.5\linewidth]{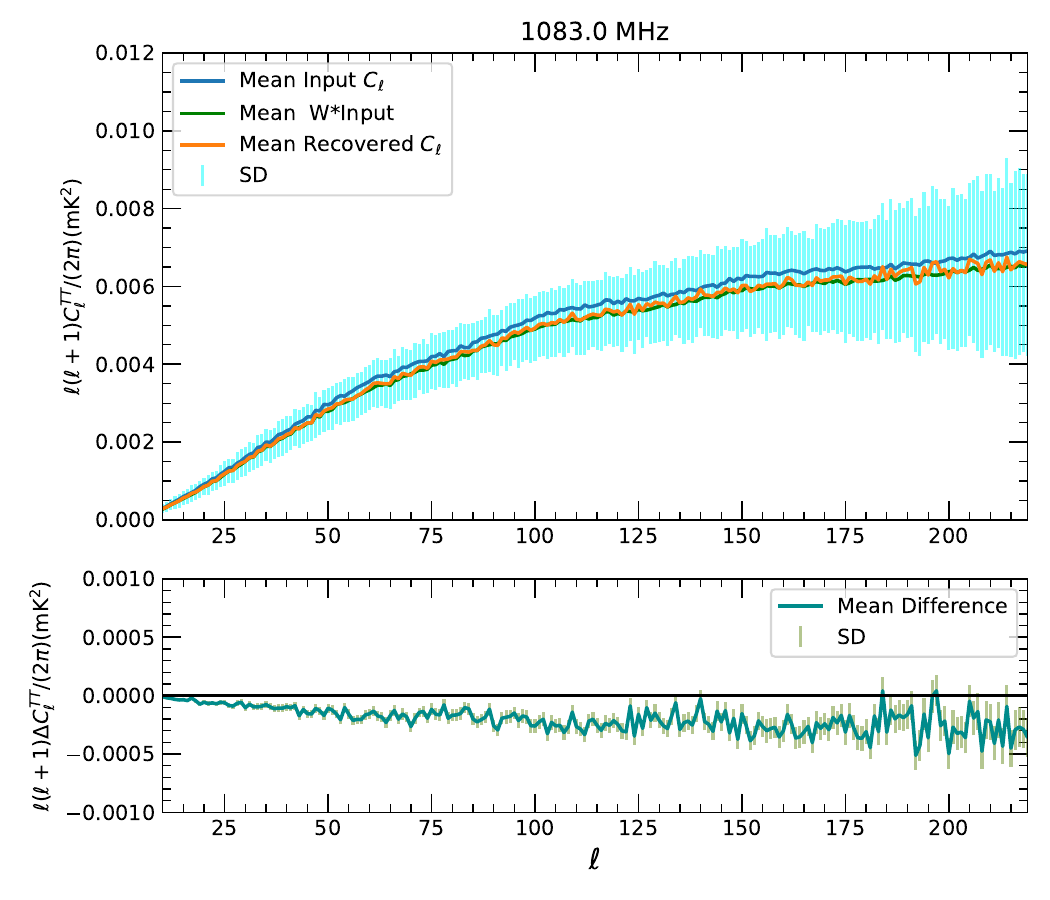}}\hfill       
	\caption{In the top left panel, we show the recovered full-sky 21 cm signal angular power spectrum corresponding to the central frequency channel 1083.0 MHz in orange. The input angular power spectrum is shown in blue. In the bottom left panel, we illustrate the difference between the recovered angular power spectrum and the input 21 cm $C_\ell^{TT}$. In the top right panel, the mean recovered full-sky angular power spectrum obtained from 200 simulations is shown in orange, along with the $1 \sigma$ standard error bars. The mean input angular power spectrum is plotted in blue, while the expected angular power spectrum recovery is shown in green. In the bottom right panel, we depict the difference between the mean recovered angular power spectrum and the mean input 21 cm signal power spectrum with $1 \sigma$ error bars. The error bars have been divided by a factor of $\sqrt{200}$. 
	}\label{fig:channel_30}
\end{figure*}
In the top left panel of Figure~\ref{fig:maps}, we show the input frequency maps comprising all foregrounds, the 21 cm signal, and noise at the central frequency channel 1209.0 MHz. The input pure 21 cm signal is shown in the top right panel. From the figures in the top  panel, it is interesting to note the several orders of magnitude differences in pixel values of the contaminated (or observed) and pure 21 cm signal. This highlights the major challenge to reconstruct the weak 21 cm signal from the strong contaminations. The bottom left panel represents the recovered 21 cm signal at the same frequency channel as mentioned above. The reconstructed cleaned map not only has comparable amplitude but also exhibits similar morphological patterns to the input pure 21 cm signal. The bottom right panel depicts the residual contamination, which is the difference between the recovered and the input 21 cm signal. The difference maps contain thermal noise and some residual foreground contaminations (the signature of residual foreground contaminations along the galactic plane may be concluded along the galactic plane after a close inspection by the eye). However, from the difference maps, it is evident that the harmonic space ILC method can effectively recover the 21 cm signal from foregrounds that are several orders of magnitude greater in intensity, with minimal residual contamination.

In Figure \ref{fig:cartview} we show the input 21 cm map and the corresponding recovered signal covering a $50^{\circ} \times 50^{\circ}$ patch of the sky at various sky positions for the 1209 MHz channel in the top and bottom panels, respectively. From left to right, these figures represent patches of the sky in the longitude and latitude ranges $[(0^{\circ}:50^{\circ}), (30^{\circ}:80^{\circ})]$, $[(200^{\circ}:250^{\circ}), (-30^{\circ}:20^{\circ})]$, $[(100^{\circ}:150^{\circ}), (-60^{\circ}:-10^{\circ})]$, and $[(250^{\circ}:300^{\circ}), (-80^{\circ}:-30^{\circ})]$, respectively. The noticeable similarity between the input and the recovered 21 cm maps in these regions suggests that our method can efficiently reconstruct the signal at various sky locations. 

 To investigate possible foreground  residuals in the cleaned maps, we construct mean different maps between the cleaned and the corresponding pure 21 cm signal at the three observed frequencies. These difference maps are shown in the top left panel of~\ref{fig:sdmaps_full}. The presence of somewhat diffused foreground residuals may be detected in the Galactic plane of all the difference maps. This is because the Galactic plane is strongly contaminated by the synchrotron and the free-free emissions. In the bottom left panel of Figure~\ref{fig:sdmaps_full}, we show the standard deviation maps corresponding to a difference map of cleaned and pure 21 cm signal obtained from the Monte Carlo simulations of foreground removal at the three different frequencies. Unlike the mean difference maps, the standard deviation maps are dominated by stronger small-scale features away from the galactic plane, indicating these standard deviation maps are dominated by the thermal noise error. The galactic plane of the standard deviation map shows clear features of residual foreground contaminations. The galactic plane, therefore, is dominated by the foreground residual errors. This suggests that the regions away from the Galactic plane are better foreground-cleaned than the Galactic plane itself.

Based upon the conclusions from Figure~\ref{fig:sdmaps_full} (left panel), we choose to excise the foreground residual regions by applying a parallel slab mask symmetrically around the galactic plane with a total width of $20^{\circ}$. We also remove bright point source locations above the galactic plane and strong point sources in the northern hemisphere. The resulting mask is shown by grey pixel regions in the right panel of Figure~\ref{fig:sdmaps_full}. The top right panel shows the same maps as the top left panel but after applying the mask. The masked maps show significant improvement in pixel values compared to the full sky cases, with maximum and minimum values occurring more symmetrically. These strongly suggest that, indeed, we are able to remove potential residual foreground-contaminated regions after masking.  The standard deviation maps in the bottom right panel, obtained by masking the corresponding bottom left panel maps, are dominated by small-scale noise features. Notably, while the minimum pixel values in both masked and unmasked standard deviation maps are identical - indicating noise dominance in low standard deviation regions - the maximum pixel values in the masked maps are somewhat smaller than their unmasked counterparts. This suggests that high standard deviation regions are dominated by residual foreground error.

\begin{figure*}[!htbp]
	\subfloat{\includegraphics[width=.5\linewidth]{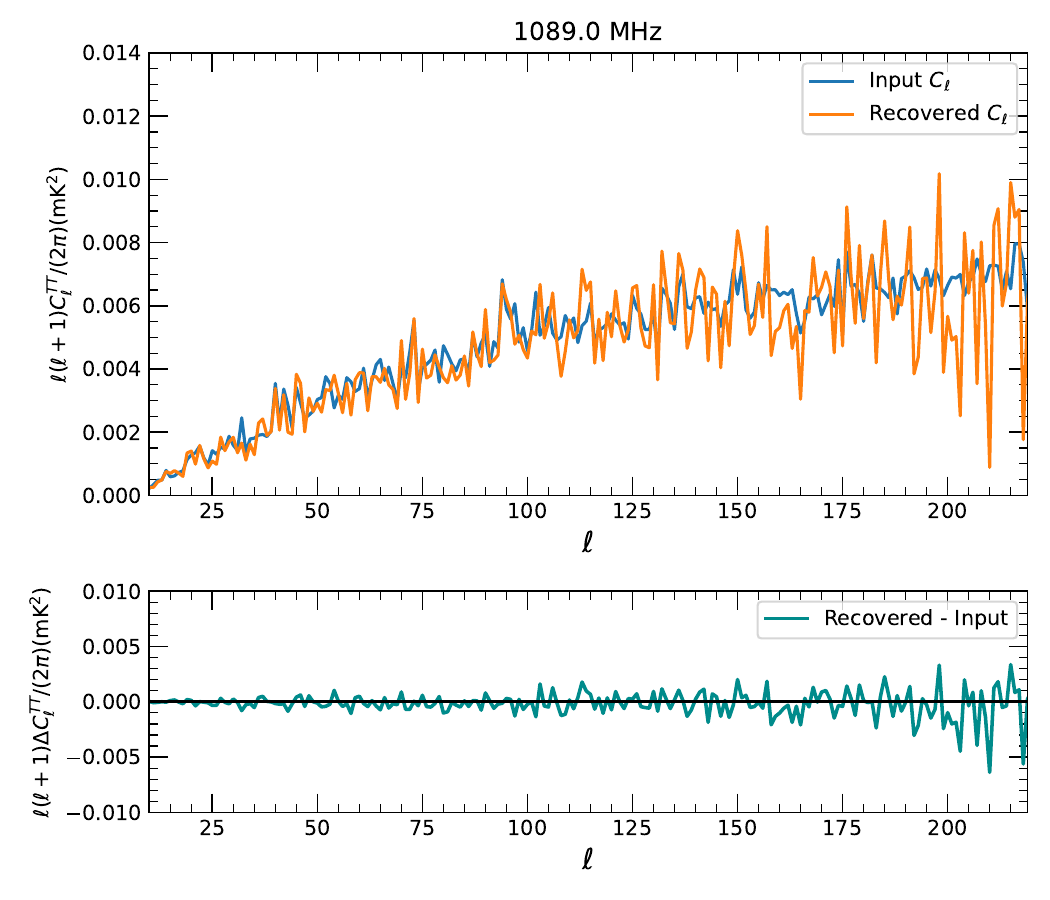}}\hfill
	\subfloat{\includegraphics[width=.5\linewidth]{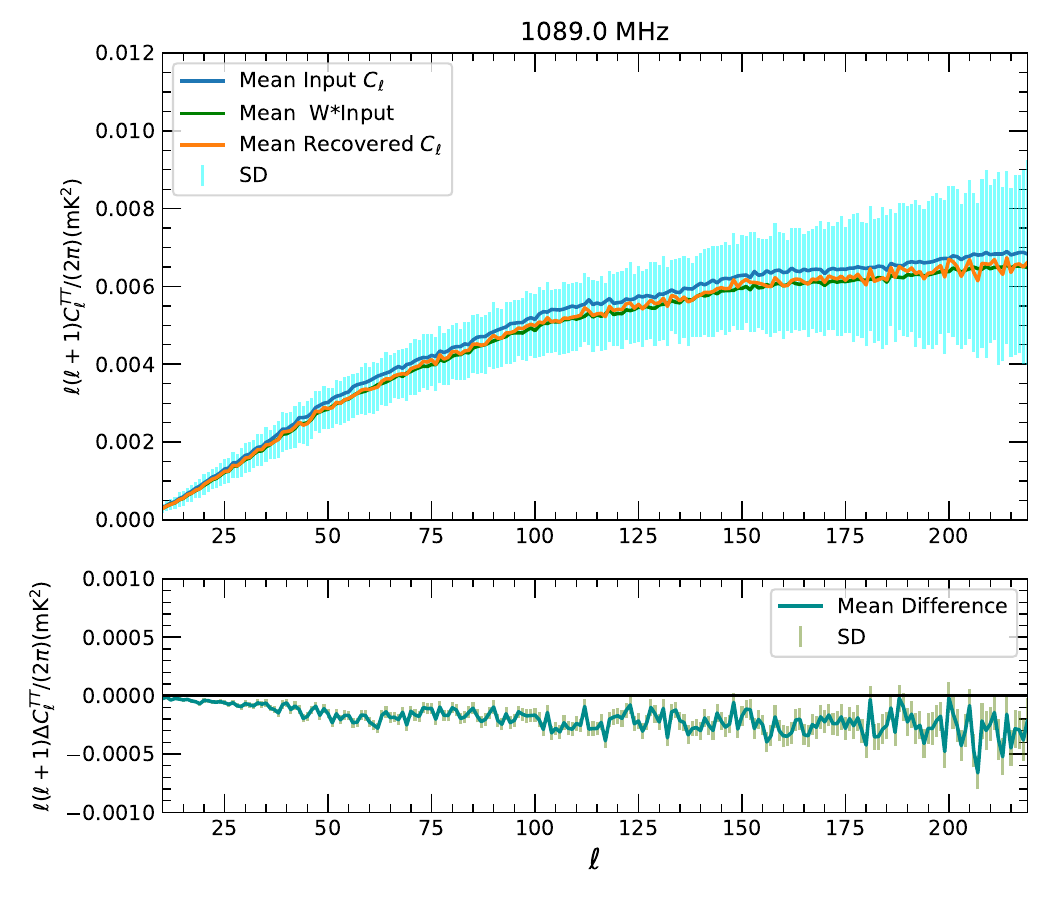}}\hfill       
	\caption{Same as Figure.~\ref{fig:channel_30} but for the recovered 21 cm full-sky angular power spectrum corresponding to the central frequency channel 1089.0 MHz. The mean absolute power reconstruction error over the simulations at each multipole is $< 5 \times 10^{-4},\text{mK}^2$. The small absolute reconstruction error at each multipole indicates that our method performs exceptionally well in reconstructing the 21 cm angular power spectrum.
	}\label{fig:channel_32}
\end{figure*}

In the top panel of Figure~\ref{fig:var_scatter}, we show scatter plots of the variances of the input pure 21 cm full-sky maps (horizontal axis) and the corresponding recovered 21 cm signals for 200 simulations at various observed frequency channels (vertical axis). Although the cleaned map variances are always larger than the input pure 21 cm map variance at any frequency shown here, these plots provide a visual representation of the correlation between the pure and cleaned 21 cm signal at a given frequency. Results for the masked maps are shown in the bottom panel of the same figure. The numerical values of the correlation coefficients between the cleaned and input pure 21 cm signal at different frequencies for the top panel are 0.76, 0.70, and 0.81 for the three frequency channels 1083 MHz, 1089 MHz, and 1209 MHz, respectively. For the bottom panel (obtained after applying the mask), the correlation coefficients change to 0.78, 0.69, and 0.83, respectively. Furthermore, in Figure~\ref{fig:corr_scatter}, we show the correlation coefficients between the input pure 21 cm maps and the corresponding recovered 21 cm signals for 200 simulations at various observed frequency channels. The input and recovered 21 cm signals are correlated at approximately $73\%, 73.5\%, \text{and} \,77.5\%$ for both full sky and masked sky at the central frequency channels of 1083.0, 1089.0, and 1209.0 MHz, respectively. Due to the statistically isotropic nature of the 21 cm signal, the correlations between the cleaned 21 cm signal and corresponding input 21 cm frequency map takes similar value for both full sky and partial sky. The correlation values are actually less than one since the pure 21 cm signal is uncorrelated between different frequencies to a good approximation and the ILC method linearly combines many input frequency maps to produce a cleaned map of any given frequency. The contributions to these linear combinations made by a different frequency channel than a chosen one cause some uncorrelation and hence a correlation coefficient less than unity.  The high correlation coefficients between the input and recovered 21 cm signals, both for the full-sky and partial-sky cases, indicate that the cleaned signal is dominantly represented by the input 21 cm signal from the same frequency channel. This also indicates that significant foreground reduction has been possible by our method.

Since our partial-sky estimates of the angular power spectrum differ from the full-sky estimates, which are directly related to the cosmological parameters, we need a methodology to convert the partial-sky estimates to the full-sky estimates of the angular power spectrum. We use the MASTER approach, specifically the mode-mode coupling matrix, to obtain the full-sky estimates of the angular power spectrum from the partial-sky power spectrum. In the following subsection, we first describe the mode-mode coupling matrix, and in the subsequent subsection, we provide our results with the full-sky angular power spectrum.

\begin{figure*}[!htbp]
	\subfloat{\includegraphics[width=.5\linewidth]{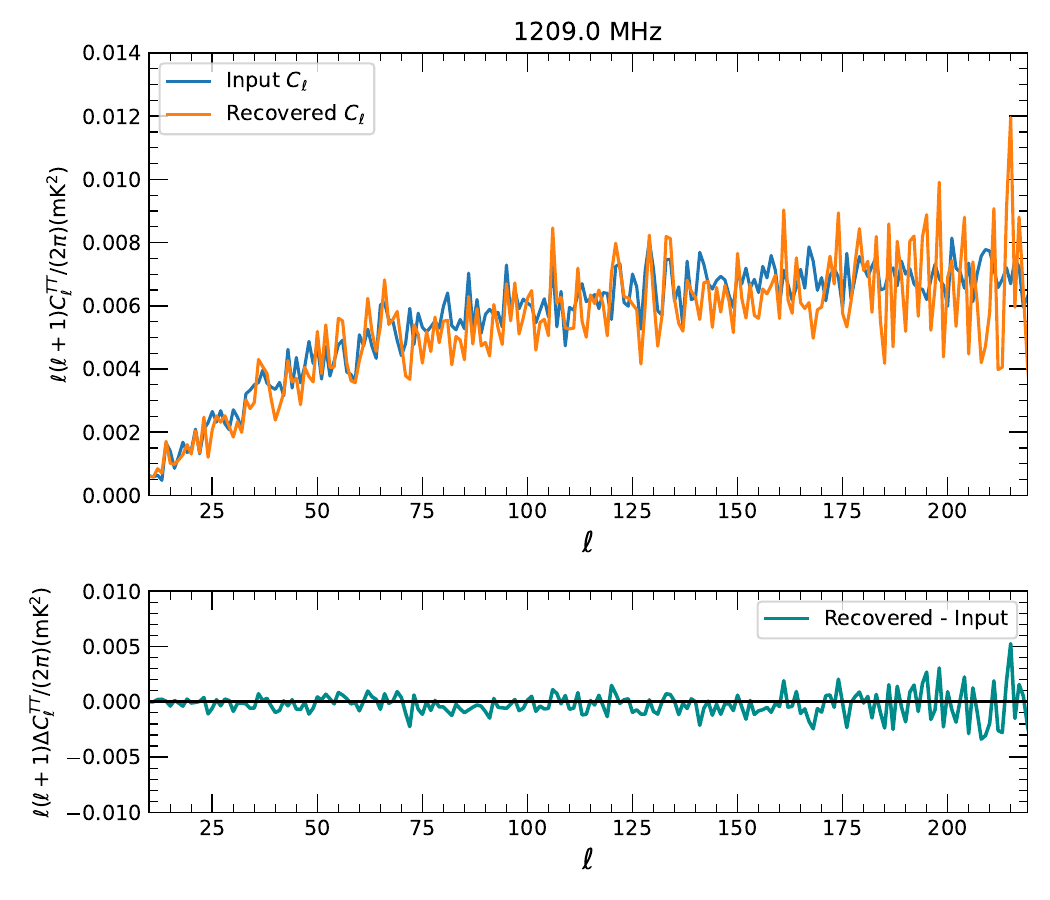}}\hfill
	\subfloat{\includegraphics[width=.5\linewidth]{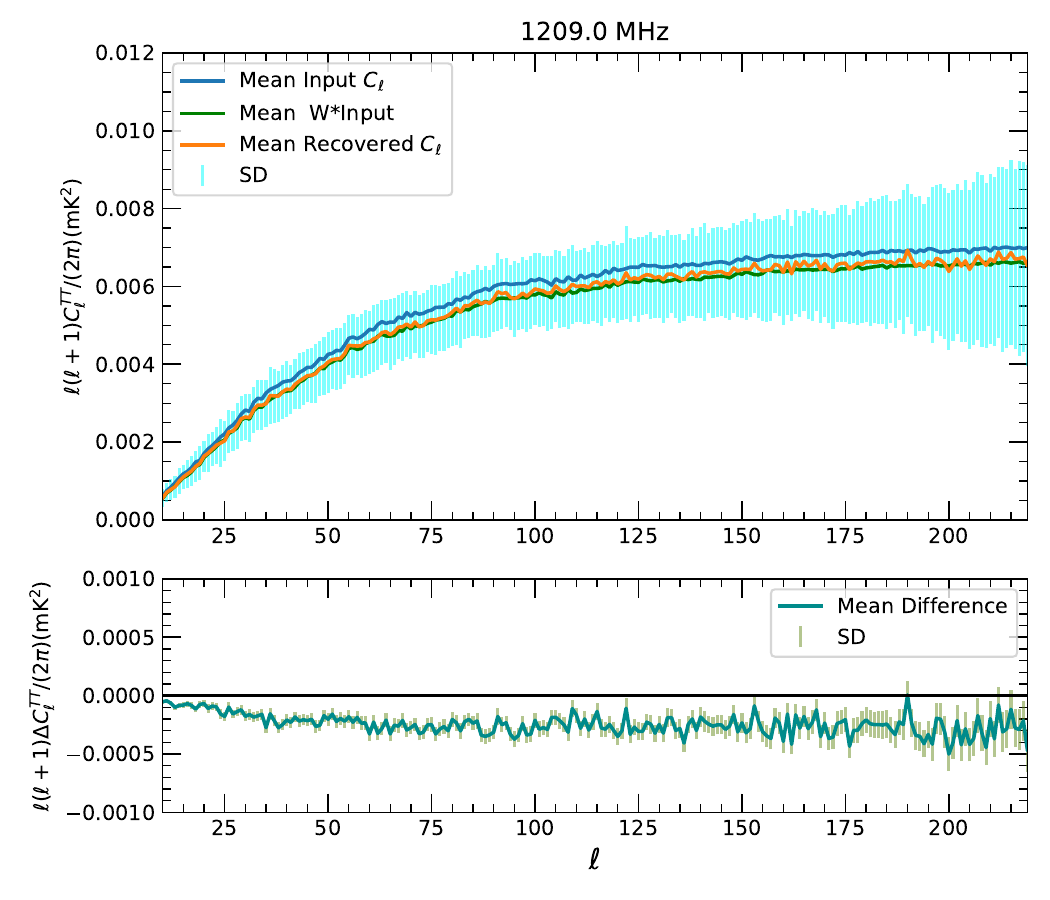}}\hfill       
	\caption{Same as Figure.~\ref{fig:channel_30} but for the recovered 21 cm full-sky angular power spectrum corresponding to the central frequency channel 1209.0 MHz. 
	}\label{fig:channel_34}
\end{figure*}

\subsection{Mode-Mode Coupling Matrix}\label{coupling_matrix}

In this subsection, we describe the MASTER (Monte Carlo Apodized Spherical Transform EstimatoR)  algorithm~\citep{2002ApJ...567....2H} implemented in our methodology which provides a powerful framework for relating the partial-sky pseudo-power spectrum $\Tilde{C_{\ell}}$, to the true full-sky power spectrum $C_{\ell}$.

The observations of the full-sky 21 cm signal $\widehat{s}({\theta, \phi})$  can be expanded in terms of spherical harmonics $Y_{\ell m}$ as,
\begin{equation}
\widehat{s}({\theta, \phi}) = \sum_{\ell > 0}\sum_{m=-\ell}^{\ell} a_{\ell m} Y_{\ell m}({\theta, \phi}) \, ,
\label{A1}
\end{equation}
where $Y_{\ell m}({\theta, \phi})$ defines spherical harmonic functions and $a_{\ell m}$ represents harmonic modes of the full sky anisotropies, which can then be written as,
\begin{eqnarray}
\qquad a_{lm} &=& \int\limits_{\theta=0}^{\pi}\int\limits_{\phi=0}^{2\pi}\widehat{s}(\theta, \phi)Y_{lm}^{*}(\theta ,\phi)d\Omega \label{a_lm}\,,
    \label{A2}
\end{eqnarray}
where $Y_{lm}^{*}(\theta,\phi)$ represents the complex conjugate of $Y_{lm}(\theta,\phi)$ and  $d\Omega$ is the elementary solid angle. Furthermore, using Equation ~\ref{A2}, we can obtain the angular power spectrum of the full-sky given by,
\begin{equation}
C_{\ell} = \frac{1}{2\ell + 1} \sum_{m=-\ell}^{\ell} \vert a_{\ell m} \vert^2 \, .
\label{A3}
\end{equation}
In the context of 21 cm observations, when the sky coverage is limited, either due to ground obscuration or galactic contamination, one can apply a position-dependent mask $M(\theta, \phi)$ to the measured data. Thus we can obtain a partial-sky map by multiplying the full-sky map with a  mask $M(\theta, \phi)$:
\begin{equation}
\tilde{s}(\theta, \phi) = { M}(\theta, \phi)  {\widehat{s}}(\theta, \phi) \, .  
\label{A4}
\end{equation}
The pseudo-harmonics $\tilde{a}_{\ell m}$ are then defined by the spherical harmonic transform of $\tilde{s}(\theta, \phi)$ i.e.,
\begin{equation}
\tilde{a}_{\ell m} = \int d\Omega \;  \tilde{s}(\theta, \phi) \; Y^{\ast}_{\ell m}(\theta, \phi)
\label{A5}
\end{equation}
The pseudo-power spectrum $\tilde{C_{\ell}}$ can then be defined as,
\begin{equation}
\tilde{C_{\ell}} = \frac{1}{2\ell + 1} \sum_{m=-\ell}^{\ell} \vert \tilde{a}_{\ell m} \vert^2 \, .
\label{A5a}
\end{equation}

Now using Equation~\ref{A4} and also by expanding  ${\widehat{s}}(\theta, \phi)$ in terms of its spherical harmonic coefficients $a_{\ell^{'}m^{'}}$, we can write Equation~\ref{A5} as,
\begin{eqnarray}
 \tilde{a}_{\ell m} &=& \sum_{\ell'm'} a_{\ell'm'} \int d\Omega  Y_{\ell'm'}(\theta, \phi) M(\theta, \phi) Y^*_{\ell m}(\theta, \phi),\\
  &=&\sum_{\ell^{'}m^{'}} a_{\ell^{'}m^{'}} K_{\ell \ell^{'}}^{mm{'}}[M] \, ,
\end{eqnarray}
where the coupling kernal  $K_{\ell \ell^{'}}^{mm{'}}$ is defined as,
\begin{equation}
K_{\ell \ell^{'}}^{mm{'}} = \int d\Omega\;{ M}(\theta, \phi)\; Y_{\ell m}(\theta, \phi)\; Y_{\ell^{'}m^{'}}^{\ast}(\theta, \phi) \, .
\label{A6}
\end{equation}

\begin{figure*}
	\centering
	\includegraphics[width=1\textwidth]{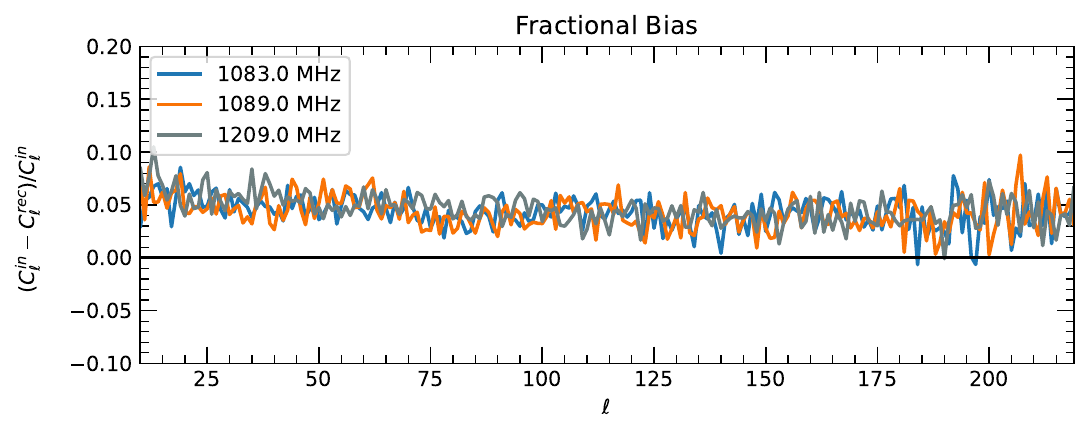}
	\caption{\small The fractional bias of the recovered 21 cm full-sky angular power spectrum at each multipole for the central frequency channels 1083.0, 1089.0, and 1209.0 MHz. The signal loss on the 21 cm power spectrum is less than $6\%$ for most multipoles, indicating that the harmonic space ILC method reconstructs the 21 cm signal with minimal loss. \normalsize
	}
	\label{fig:fbias}
\end{figure*}
This can be further expressed as,
\begin{equation}
K_{\ell \ell^{'}}^{mm{'}} = \sum_{\ell^{\prime\prime} m^{\prime\prime}} a_{\ell^{\prime\prime} m^{\prime\prime}}^M \int d\Omega Y_{\ell^{\prime\prime} m^{\prime\prime}}(\theta, \phi) Y_{\ell^\prime m^\prime}^{\ast}(\theta, \phi)Y_{\ell m}(\theta, \phi)\,  
\label{A7}
\end{equation}
\begin{equation}
= \sum_{\ell^{\prime\prime} m^{\prime\prime}} a_{\ell^{\prime\prime} m^{\prime\prime}}^{M} (-1)^{m^{\prime}} \Big[ \frac{(2\ell^{\prime\prime} +1)(2\ell^{\prime} + 1)(2\ell+1)}{4\pi} \Big]^{1/2}
\end{equation}
\begin{equation*}
\begin{pmatrix}
\ell & \ell^{\prime} & \ell^{\prime\prime} \\ 0 & 0 & 0
\end{pmatrix}
\begin{pmatrix}
\ell & \ell^{\prime} & \ell^{\prime\prime} \\ m & -m^{\prime} & m^{\prime\prime}
\end{pmatrix} \, .
\end{equation*}
with the spherical harmonic transform coefficient of the mask given by,
\begin{equation}
a_{\ell m}^{M} = \int  d\Omega \; {M}(\theta, \phi) \; Y^{\ast}_{\ell m}(\theta, \phi) \, .
\label{A8}
\end{equation}
Here, $\begin{pmatrix}
\ell & \ell^{\prime} & \ell^{\prime\prime} \\ m & m^{\prime} & m^{\prime\prime}
\end{pmatrix}$ is the Wigner 3j symbol, which is related to the Clebsch-Gordan coefficients and describes the coupling of angular momenta. They are non-zero only if they satisfy selection rules,
\begin{equation}
        m + m^{\prime} + m^{\prime\prime} = 0,
        \label{A9}
\end{equation}
and,
\begin{equation}
    |\ell-\ell^{\prime}| \leq  \ell^{\prime\prime} \leq \ell+\ell^{\prime}\,.
    \label{A10}
\end{equation}
The orthogonality relations of the Weigner symbols reads,
\begin{equation}
\sum_{\ell^{\prime\prime} m^{\prime\prime}} (2\ell^{\prime\prime} + 1) \begin{pmatrix}
\ell & \ell^{\prime} & \ell^{\prime\prime} \\ m & m^{\prime} & m^{\prime\prime}
\end{pmatrix}
\begin{pmatrix}
\ell & \ell^{\prime} & \ell^{\prime\prime} \\ m^\ast & m^{\ast\prime} & m^{\prime\prime}
\end{pmatrix} = \delta_{mm^\ast}\delta_{m^\prime m^{\ast\prime}} \, ,
\end{equation}
\begin{multline}
\sum_{m m^{\prime}} \begin{pmatrix}
\ell & \ell^{\prime} & \ell^{\prime\prime} \\ m & m^{\prime} & m^{\prime\prime}
\end{pmatrix}
\begin{pmatrix}
\ell & \ell^{\prime} & \ell^{\ast\prime\prime} \\ m & m^{\prime} & m^{\ast\prime\prime}
\end{pmatrix} = \delta_{\ell^{\prime\prime}\ell{^\ast\prime\prime}}\delta_{m^{\prime\prime} m^{\ast\prime\prime}}  \\ \delta(\ell, \ell^\prime, \ell^{\prime\prime})\frac{1}{2\ell^{\prime\prime}+1} \, ,
\end{multline}
where $\delta(\ell, \ell^\prime, \ell^{\prime\prime}) = 1$ when Equation~\ref{A10} is satisfied else $\delta(\ell, \ell^\prime, \ell^{\prime\prime}) = 0$.

Substituting the expression for the coupling kernel into the pseudo-power spectrum and using the orthogonality properties of the Wigner 3j-symbols, we get:
\begin{equation}
\tilde{C}_{\ell} = M_{\ell\ell^\prime}C_{\ell^{\prime}} \,  . 
\label{partcl}
\end{equation}
where the mode-mode coupling matrix $M_{\ell\ell^\prime}$ is given by:
\begin{equation}
M_{\ell\ell^\prime} = \frac{2\ell^\prime + 1}{4\pi} \sum_{\ell^{\prime\prime}} (2\ell^{\prime\prime} + 1) C_{\ell^{\prime\prime}}^{M} \begin{pmatrix}
\ell & \ell^\prime & \ell^{\prime\prime} \\ 0 & 0 & 0
\end{pmatrix}^{2}\, ,
\label{mllp}
\end{equation}
and $C_\ell^M$ is the power spectrum of the smoothed mask $M$.
If the sky cut is small, then the $M_{\ell\ell^\prime}$ matrix appearing in Equation~\ref{mllp} will be non-singular and hence invertible. So to recover the full-sky power spectrum $C_\ell$ from the partial-sky pseudo-spectrum $\tilde{C}_{\ell^{\prime}}$, one needs to invert the coupling matrix:
\begin{equation}
{C}_{\ell} = M_{\ell\ell^\prime}^{-1}\tilde{C}_{\ell^{\prime}}  \,  . 
\label{reccl}
\end{equation}

\subsection{Full-sky Angular Power Spectrum}\label{aps}

In this subsection, we present our method's ability to reconstruct the 21 cm full-sky angular power spectrum in the multipole range $\ell < 220$ by inverting the mode-mode coupling matrix following Equation~\ref{reccl}. The Figure~\ref{fig:cl_all} shows the power spectra of various astrophysical sources of contamination and thermal noise in comparison with the theoretical 21 cm power spectrum for the central frequency channel 1083.0 MHz. The synchrotron emission (orange), free-free emission (red) and extragalactic point sources (blue) dominate the observed sky signal (green) by several orders of magnitude compared to the 21 cm signal (purple). This stark difference in the power spectra highlights the major challenge in detecting the weak 21 cm signal amidst the overwhelmingly dominant foreground contamination. So here we address this challenge by demonstrating our methodology's ability to reconstruct the 21 cm angular power spectrum accurately.
\begin{figure}
	\centering
	\includegraphics[width=0.5\textwidth]{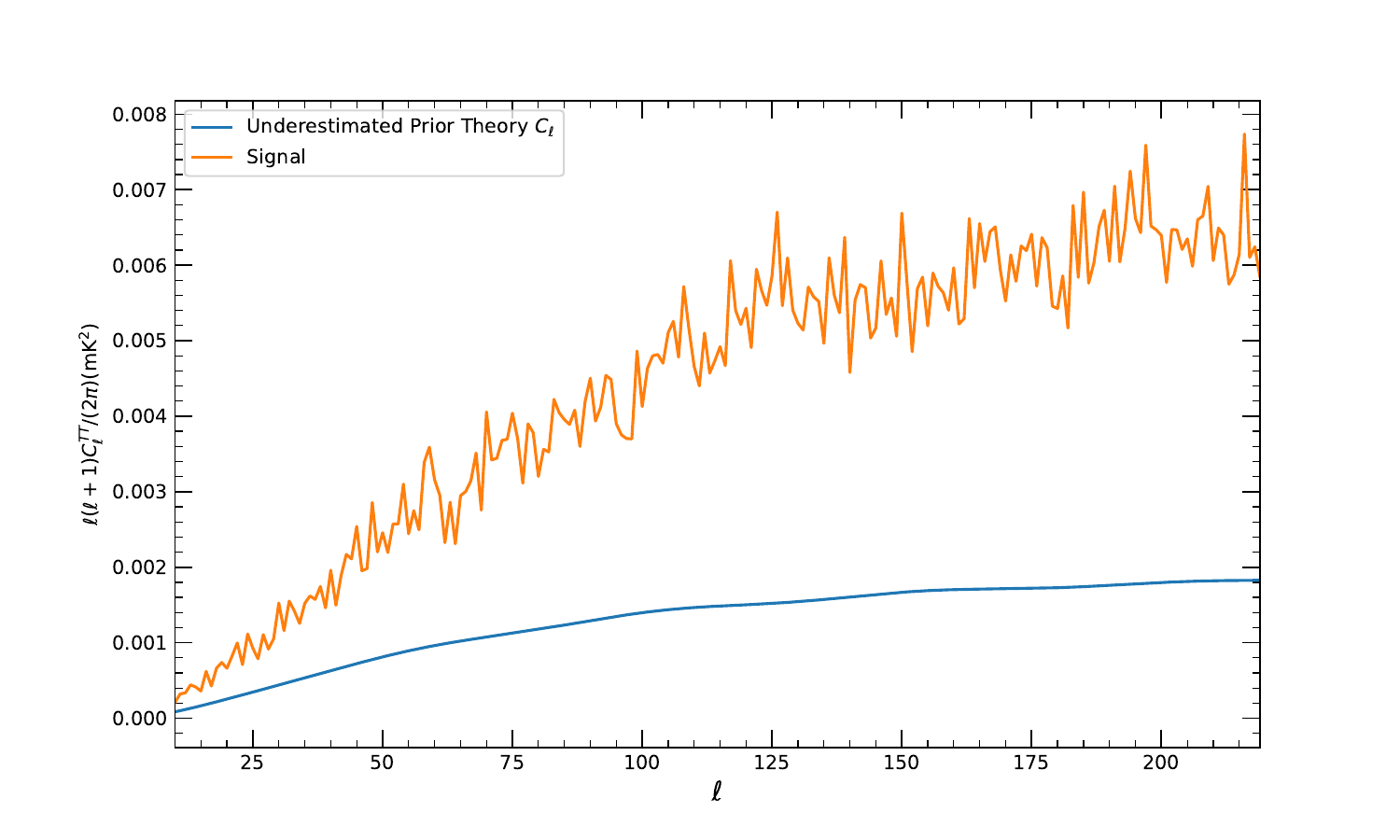}
	\caption{\small Comparison of angular power spectra showing the magnitude of prior covariance underestimation at central frequency channel 1089 MHz. This significant difference between the blue and orange curve is due to large deviations ($2\sigma$) in the neutral hydrogen density.  \normalsize
	}
	\label{fig:underestimate}
\end{figure}
The full-sky angular power spectrum of the reconstructed 21 cm signal corresponding to the central frequency channel 1089.0 MHz is shown in the top left panel of Figure~\ref{fig:channel_30}. The recovered 21 cm signal power spectrum is plotted in orange, while the input power spectrum is depicted in blue. The recovered 21 cm full-sky power spectrum aligns well with the input 21 cm spectrum. The difference full-sky angular power spectrum obtained from subtracting the input 21 cm power spectrum from the recovered power spectrum is shown in the bottom left panel of Figure~\ref{fig:channel_30}. In the top right panel of Figure~\ref{fig:channel_30}, we show the mean over 200 simulations of the input and recovered angular power spectra along with the corresponding standard deviations to quantify the reconstruction error in the 21 cm angular power spectrum. In the bottom right panel of Figure~\ref{fig:channel_30}, we plot the mean over 200 simulations of the difference between the recovered and input angular power spectra along with the corresponding standard deviations to further quantify the reconstruction error in the recovered 21 cm signal angular power spectrum. 
Furthermore, in Figure~\ref{fig:channel_32} and Figure~\ref{fig:channel_34}, we also show the reconstruction of the 21 cm angular power spectrum corresponding to the central frequency channels 1089.0 MHz and 1209.0 MHz, respectively, indicating that our method can reconstruct the 21 cm power spectrum over a wide range of frequency channels. In Figure~\ref{fig:fbias} we show the fractional bias of the recovered 21 cm full-sky angular power spectrum at each multipole for the central frequency channels 1083.0, 1089.0, and 1209.0 MHz. The signal loss on the 21 cm power spectrum is less than $6\%$ for most multipoles, indicating that the harmonic space ILC method reconstructs the 21 cm signal with minimal loss.

\begin{figure*}
	\centering
	\includegraphics[width=1\textwidth]{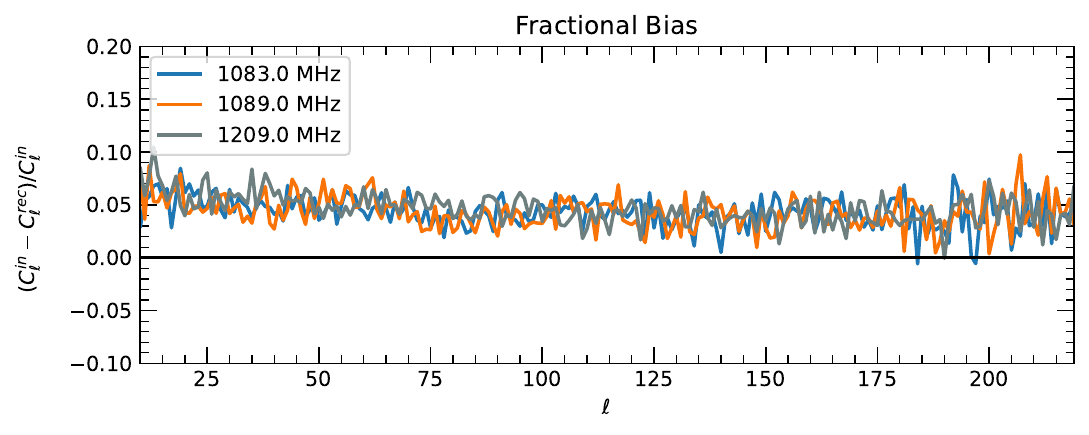}
	\caption{\small The fractional bias of the recovered 21 cm full-sky angular power spectrum at each multipole for the central frequency channels 1083.0, 1089.0, and 1209.0 MHz, when the prior theoretical covariance underestimates the signal. \normalsize
	}
	\label{fig:fbias_cov}
\end{figure*}

\section{Influence of unknown systematics}\label{systematics}
In real observations, various systematic effects can impact the effectiveness of foreground removal techniques. Here we investigate the robustness of our harmonic space ILC method against two important systematic effects: bias in the theoretical prior covariance and residual gain errors.

\subsection{Effect of Prior Covariance Underestimation}

A critical assumption in our methodology is the use of theoretical prior covariance for the 21 cm signal.  To rigorously assess our method's robustness against inaccurate prior assumptions, we analyze cases where the assumed theoretical covariance systematically underestimates the true signal covariance by $2\sigma$, by considering a neutral hydrogen density  $\Omega_{\rm HI}$ that is two standard deviations below the measured  value constrained by~\cite{2013MNRAS.434L..46S}. This substantial deviation represents potential systematic uncertainties that could arise from theoretical modeling limitations or imprecise cosmological parameter constraints in a moderately  pessimistic scenario. We show the prior theoretical power spectrum used for this analyses against the true signal power spectrum in Figure~\ref{fig:underestimate}, illustrating the magnitude of the underestimation at the chosen frequency of 1089MHz. Through detailed Monte Carlo simulations with 200 realizations, we find that even with this substantial difference between assumed and actual covariance, the results are consistent with our baseline results. Figure~\ref{fig:fbias_cov} shows the fractional bias in the recovered 21 cm power spectrum under these imperfect prior assumptions. The results demonstrate that our method maintains its effectiveness, with the fractional bias remaining below 7\% in the mid multipole range $\ell \sim 70-180$. 

\subsection{Effect of Residual Gain Error}

\begin{figure*}
	\centering
	\includegraphics[width=1\textwidth]{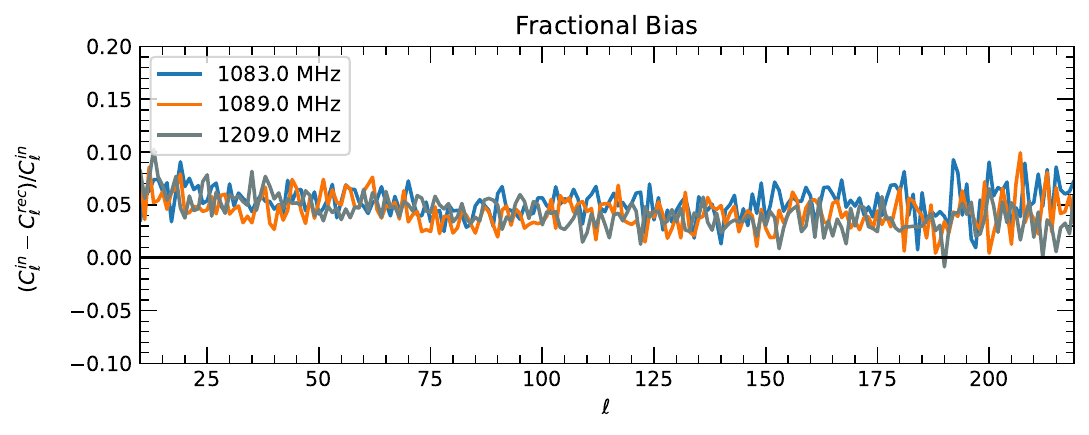}
	\caption{\small The fractional bias of the recovered 21 cm full-sky angular power spectrum at each multipole for the central frequency channels 1083.0, 1089.0, and 1209.0 MHz, after incorporating 2\% residual gain errors.  \normalsize
	}
	\label{fig:fbias_gain}
\end{figure*}

Gain calibration errors represent another important systematic effect in radio observations. To assess our method's resilience to such errors, we introduce a 2\% residual gain error scaled with the input total signal before processing through our pipeline. This level of gain error is significantly higher than typical uncertainties in current generation experiments, providing a siginificant test of the robustness of our method in a sufficiently  difficult scenario. Our analysis shows that the harmonic space ILC method handles these gain errors effectively. As shown in Figure~\ref{fig:fbias_gain}, the additional fractional bias introduced by the 2\% gain error remains small and is comparable to our baseline results, maintaining a fractional bias below $7\%$ across mid multipole ranges. This demonstrates that our methodology is robust against realistic levels of gain calibration uncertainty.

These results suggest that our method can maintain its effectiveness under realistic observing conditions and in the presence of significant uncertainties in theoretical modeling of the 21 cm signal, making it a promising tool for upcoming 21 cm intensity mapping experiments.
\\
\\
\section{DISCUSSION AND CONCLUSIONS}\label{conclusions}
The accurate estimation of the 21 cm signal provides valuable insights into previously unexplored stages of the Universe, spanning the Dark Ages, the Cosmic Dawn, and post-reionization phases. Since neutral hydrogen were ubiquitously present in the Universe starting from the epoch of recombination and modified its spatial distribution thereafter due to structure formation, 21 cm signal  from neutral hydrogen uniquely encodes the information about the structure formation in the Universe. The observations of the large-scale clustering of matter hold significant promise for advancing our understanding of both early and late universe phenomena, including cosmic inflation, dark energy, and galaxy evolution~\citep{2020PASP..132f2001L,2023MNRAS.519.1809J,2022MNRAS.511.1637J,2021JApA...42..111J}.
In this study, we propose a novel, foreground model-independent method for extracting the 21 cm signal  using simulated observations of future generation single dish observation BINGO by removing  dominant foreground components. We introduce the internal linear combination in harmonic space, leveraging prior knowledge of the theoretical 21 cm covariance matrix for accurate estimation of the 21 cm signal and its angular power spectrum. The harmonic space ILC maximizes the variance of the 21 cm signal by linearly combining the frequency maps using certain weights that filter out exactly the same subspace as spanned by the linearly independent templates of the 21cm signal. We extensively test the ability of our method to reconstruct the 21 cm signal and its power spectrum over the BINGO bandwidth, which corresponds to a redshift range of 0.13 to 0.48. Our study employs the simulated radio emissions incorporating the synchrotron emission, free-free emission and the extragalactic point sources. Regarding the instrument noise, we consider the thermal noise.

We summarize the ﬁndings of our method as follows.
\begin{enumerate}
\item The ILC method in combination with a PCA technique in the harmonic space one can reconstruct the 21 cm signal from the observed maps that are contaminated by the foregrounds (and detector noise) that are several orders of magnitude larger than the signal.

\item Our method is particularly very useful since the foregrounds are much stronger than the signal. For methods that use explicit foreground modelling to remove the foregrounds may be prone to foreground leakage in the final estimated signal due to any incorrect foreground modelling. The problem may be  more severe for weak 21 cm signal since a small modelling error in strong foregrounds  can lead to large contamination in the reconstructed signal. The ILC method avoids such problem and can lead to accurate foreground removal in the limit when sufficient observation frequencies are available without any need to explicitly model the foregrounds. 

\item We utilize $50$ different frequency channels from the future generation BINGO single telescope observation as well as implement the ILC method in the harmonic space and produce full sky foreground minimized 21 cm maps at the central frequencies of BINGO. Using the large number of frequency channels helps removing the synchrotron emissions with varying spectral index effectively. Also working in the harmonic space provides  a method to incorporate variation of spectral parameters with scales. 

\item Along with the foreground cleaned maps we also estimate the angular power spectrum of 21 cm signal after first applying a galactic mask. Since the full sky angular spectrum of the 21 cm signal can be used to directly measure the cosmological parameters we provide estimates of the full sky 21 cm power spectrum from the masked sky by using mode-mode coupling matrix~\citep{2002ApJ...567....2H}. Based upon detailed Monte Carlo simulations of the entire  foreground  removal and power spectrum pipeline we show that the reconstructed map and partial sky angular spectra agree well with their underlying counterparts.

\item Our method is computationally fast since we work in harmonic space. The entire foreground removal and power spectrum estimation for a given observation realization takes only around two minutes with an Intel\textsuperscript{®} Core\texttrademark{} i5-9300H CPU @ 2.40GHz processor and 4 cores. This allows us to perform detailed Monte Carlo estimations of foreground cleaned 21 cm map and its full sky power spectrum estimate using Mode-mode coupling matrix. Using the mean recovered spectra after correcting for the mode coupling we find the residual negative bias in the 21 cm cleaned spectra. We note that this bias is entirely expected since principal components are employed. The bias estimates obtained from the Monte Carlo simulations can be applied during the analysis with the observed data. The bias corrected spectra can then be used for cosmological parameter estimation. 

\item We demonstrate our method's robustness against key systematic effects through  extensive validation. The method maintains its effectiveness when challenged with both theoretical prior misspecification ($2\sigma$ underestimation of signal covariance) and instrumental uncertainties ($2\%$ residual gain errors). In both scenarios, the fractional bias remains below $7\% $ across mid multipole ranges ($\ell$ $\approx$ 70-180), validating the method's reliability under realistic observational conditions and theoretical uncertainties.

\end{enumerate}

Though in our analysis we consider only 50 frequency channels, the 21 cm intensity mapping experiments like BINGO will have approximately 1000 frequency channels in the 960 MHz to 1260 MHz band. In such real scenarios, the effective number of independent parameters combining all foreground components will be significantly smaller than the total number of frequency maps. This makes the ILC method, in combination with PCA method, perform excellently well in removing the foreground, which effectively contains fewer independent parameters than the available frequency maps. The exceptional performance of our method underscores its potential for intensity mapping experiments, offering a promising avenue for future studies in 21 cm cosmology. While we perform analysis at low redshifts, this technique can also be applied at higher redshifts to probe the Epoch of Reionization.\\
\
\\

\section*{ACKNOWLEDGEMENTS}
AJ thanks Simon Foreman for enlightening discussions associated with this work. We use publicly available HEALPix~\citep{2005ApJ...622..759G} package available from http://healpix.sourceforge.net for  the analysis of this work.

\bibliography{ms}{}

\begin{thebibliography}{}
\expandafter\ifx\csname natexlab\endcsname\relax\def\natexlab#1{#1}\fi
\providecommand{\url}[1]{\href{#1}{#1}}
\providecommand{\dodoi}[1]{doi:~\href{http://doi.org/#1}{\nolinkurl{#1}}}
\providecommand{\doeprint}[1]{\href{http://ascl.net/#1}{\nolinkurl{http://ascl.net/#1}}}
\providecommand{\doarXiv}[1]{\href{https://arxiv.org/abs/#1}{\nolinkurl{https://arxiv.org/abs/#1}}}

\bibitem[{{Alonso} {et~al.}(2015){Alonso}, {Bull}, {Ferreira}, \&
  {Santos}}]{2015MNRAS.447..400A}
{Alonso}, D., {Bull}, P., {Ferreira}, P.~G., \& {Santos}, M.~G. 2015, \mnras,
  447, 400, \dodoi{10.1093/mnras/stu2474}

\bibitem[{{Amiri} {et~al.}(2023){Amiri}, {Bandura}, {Chen}, {Deng}, {Dobbs},
  {Fandino}, {Foreman}, {Halpern}, {Hill}, {Hinshaw}, {H{\"o}fer}, {Kania},
  {Landecker}, {MacEachern}, {Masui}, {Mena-Parra}, {Milutinovic},
  {Mirhosseini}, {Newburgh}, {Ordog}, {Pen}, {Pinsonneault-Marotte}, {Polzin},
  {Reda}, {Renard}, {Shaw}, {Siegel}, {Singh}, {Vanderlinde}, {Wang}, {Wiebe},
  {Wulf}, \& {CHIME Collaboration}}]{2023ApJ...947...16A}
{Amiri}, M., {Bandura}, K., {Chen}, T., {et~al.} 2023, \apj, 947, 16,
  \dodoi{10.3847/1538-4357/acb13f}

\bibitem[{{Barkana}(2018)}]{2018Natur.555...71B}
{Barkana}, R. 2018, \nat, 555, 71, \dodoi{10.1038/nature25791}

\bibitem[{Basak \& Delabrouille(2011)}]{10.1111/j.1365-2966.2011.19770.x}
Basak, S., \& Delabrouille, J. 2011, Monthly Notices of the Royal Astronomical
  Society, 419, 1163, \dodoi{10.1111/j.1365-2966.2011.19770.x}

\bibitem[{{Battye} {et~al.}(2013){Battye}, {Browne}, {Dickinson}, {Heron},
  {Maffei}, \& {Pourtsidou}}]{2013MNRAS.434.1239B}
{Battye}, R.~A., {Browne}, I.~W.~A., {Dickinson}, C., {et~al.} 2013, \mnras,
  434, 1239, \dodoi{10.1093/mnras/stt1082}

\bibitem[{{Battye} {et~al.}(2004){Battye}, {Davies}, \&
  {Weller}}]{2004MNRAS.355.1339B}
{Battye}, R.~A., {Davies}, R.~D., \& {Weller}, J. 2004, \mnras, 355, 1339,
  \dodoi{10.1111/j.1365-2966.2004.08416.x}

\bibitem[{{Bedini} {et~al.}(2005){Bedini}, {Herranz}, {Salerno}, {Baccigalupi},
  {Kuruouglu}, \& {Tonazzini}}]{2005EJASP2005.2400B}
{Bedini}, L., {Herranz}, D., {Salerno}, E., {et~al.} 2005, EURASIP Journal on
  Applied Signal Processing, 2005, 2400, \dodoi{10.1155/ASP.2005.2400}

\bibitem[{{Bharadwaj} {et~al.}(2001){Bharadwaj}, {Nath}, \&
  {Sethi}}]{2001JApA...22...21B}
{Bharadwaj}, S., {Nath}, B.~B., \& {Sethi}, S.~K. 2001, Journal of Astrophysics
  and Astronomy, 22, 21, \dodoi{10.1007/BF02933588}

\bibitem[{{Bharadwaj} \& {Sethi}(2001)}]{2001JApA...22..293B}
{Bharadwaj}, S., \& {Sethi}, S.~K. 2001, Journal of Astrophysics and Astronomy,
  22, 293, \dodoi{10.1007/BF02702273}

\bibitem[{{Bottino} {et~al.}(2010){Bottino}, {Banday}, \&
  {Maino}}]{2010MNRAS.402..207B}
{Bottino}, M., {Banday}, A.~J., \& {Maino}, D. 2010, \mnras, 402, 207,
  \dodoi{10.1111/j.1365-2966.2009.15917.x}

\bibitem[{{Bouchet} {et~al.}(1999){Bouchet}, {Prunet}, \&
  {Sethi}}]{1999MNRAS.302..663B}
{Bouchet}, F.~R., {Prunet}, S., \& {Sethi}, S.~K. 1999, \mnras, 302, 663,
  \dodoi{10.1046/j.1365-8711.1999.02118.x}

\bibitem[{{Bull} {et~al.}(2015){Bull}, {Ferreira}, {Patel}, \&
  {Santos}}]{2015ApJ...803...21B}
{Bull}, P., {Ferreira}, P.~G., {Patel}, P., \& {Santos}, M.~G. 2015, \apj, 803,
  21, \dodoi{10.1088/0004-637X/803/1/21}

\bibitem[{{Burba} {et~al.}(2024){Burba}, {Bull}, {Wilensky}, {Kennedy},
  {Garsden}, \& {Glasscock}}]{2024MNRAS.535..793B}
{Burba}, J., {Bull}, P., {Wilensky}, M.~J., {et~al.} 2024, \mnras, 535, 793,
  \dodoi{10.1093/mnras/stae2334}

\bibitem[{{Carucci} {et~al.}(2020){Carucci}, {Irfan}, \&
  {Bobin}}]{2020MNRAS.499..304C}
{Carucci}, I.~P., {Irfan}, M.~O., \& {Bobin}, J. 2020, \mnras, 499, 304,
  \dodoi{10.1093/mnras/staa2854}

\bibitem[{{Chang} {et~al.}(2010){Chang}, {Pen}, {Bandura}, \&
  {Peterson}}]{nature1}
{Chang}, T.-C., {Pen}, U.-L., {Bandura}, K., \& {Peterson}, J.~B. 2010, \nat,
  466, 463, \dodoi{10.1038/nature09187}

\bibitem[{{Chang} {et~al.}(2008){Chang}, {Pen}, {Peterson}, \&
  {McDonald}}]{2008PhRvL.100i1303C}
{Chang}, T.-C., {Pen}, U.-L., {Peterson}, J.~B., \& {McDonald}, P. 2008, \prl,
  100, 091303, \dodoi{10.1103/PhysRevLett.100.091303}

\bibitem[{{CHIME Collaboration} {et~al.}(2022){CHIME Collaboration}, {Amiri},
  {Bandura}, {Boskovic}, {Chen}, {Cliche}, {Deng}, {Denman}, {Dobbs},
  {Fandino}, {Foreman}, {Halpern}, {Hanna}, {Hill}, {Hinshaw}, {H{\"o}fer},
  {Kania}, {Klages}, {Landecker}, {MacEachern}, {Masui}, {Mena-Parra},
  {Milutinovic}, {Mirhosseini}, {Newburgh}, {Nitsche}, {Ordog}, {Pen},
  {Pinsonneault-Marotte}, {Polzin}, {Reda}, {Renard}, {Shaw}, {Siegel},
  {Singh}, {Smegal}, {Tretyakov}, {van Gassen}, {Vanderlinde}, {Wang}, {Wiebe},
  {Willis}, \& {Wulf}}]{2022ApJS..261...29C}
{CHIME Collaboration}, {Amiri}, M., {Bandura}, K., {et~al.} 2022, \apjs, 261,
  29, \dodoi{10.3847/1538-4365/ac6fd9}

\bibitem[{{CHIME Collaboration} {et~al.}(2023){CHIME Collaboration}, {Amiri},
  {Bandura}, {Chakraborty}, {Dobbs}, {Fandino}, {Foreman}, {Gan}, {Halpern},
  {Hill}, {Hinshaw}, {H{\"o}fer}, {Landecker}, {Li}, {MacEachern}, {Masui},
  {Mena-Parra}, {Milutinovic}, {Mirhosseini}, {Newburgh}, {Ordog}, {Paul},
  {Pen}, {Pinsonneault-Marotte}, {Reda}, {Shaw}, {Siegel}, {Vanderlinde},
  {Wang}, {Wiebe}, \& {Wulf}}]{2023arXiv230904404C}
---. 2023, arXiv e-prints, arXiv:2309.04404, \dodoi{10.48550/arXiv.2309.04404}

\bibitem[{{Chowdhury} {et~al.}(2020){Chowdhury}, {Kanekar}, {Chengalur},
  {Sethi}, \& {Dwarakanath}}]{2020Natur.586..369C}
{Chowdhury}, A., {Kanekar}, N., {Chengalur}, J.~N., {Sethi}, S., \&
  {Dwarakanath}, K.~S. 2020, \nat, 586, 369, \dodoi{10.1038/s41586-020-2794-7}

\bibitem[{{Crichton} \& et.al.(2022)}]{2022JATIS...8a1019C}
{Crichton}, D., \& et.al. 2022, Journal of Astronomical Telescopes,
  Instruments, and Systems, 8, 011019, \dodoi{10.1117/1.JATIS.8.1.011019}

\bibitem[{{Datta} {et~al.}(2007){Datta}, {Choudhury}, \&
  {Bharadwaj}}]{2007MNRAS.378..119D}
{Datta}, K.~K., {Choudhury}, T.~R., \& {Bharadwaj}, S. 2007, \mnras, 378, 119,
  \dodoi{10.1111/j.1365-2966.2007.11747.x}

\bibitem[{{Delabrouille} \& {Cardoso}(2007)}]{2007astro.ph..2198D}
{Delabrouille}, J., \& {Cardoso}, J.~F. 2007, arXiv e-prints, astro,
  \dodoi{10.48550/arXiv.astro-ph/0702198}

\bibitem[{{Di Matteo} {et~al.}(2002){Di Matteo}, {Perna}, {Abel}, \&
  {Rees}}]{2002ApJ...564..576D}
{Di Matteo}, T., {Perna}, R., {Abel}, T., \& {Rees}, M.~J. 2002, \apj, 564,
  576, \dodoi{10.1086/324293}

\bibitem[{{Dickinson}(2012)}]{2012rsri.confE..34D}
{Dickinson}, C. 2012, in Resolving The Sky - Radio Interferometry: Past,
  Present and Future, 34, \dodoi{10.22323/1.163.0034}

\bibitem[{{Drinkwater} {et~al.}(2010){Drinkwater}, {Jurek}, {Blake}, {Woods},
  {Pimbblet}, {Glazebrook}, {Sharp}, {Pracy}, {Brough}, {Colless}, {Couch},
  {Croom}, {Davis}, {Forbes}, {Forster}, {Gilbank}, {Gladders}, {Jelliffe},
  {Jones}, {Li}, {Madore}, {Martin}, {Poole}, {Small}, {Wisnioski}, {Wyder}, \&
  {Yee}}]{2010MNRAS.401.1429D}
{Drinkwater}, M.~J., {Jurek}, R.~J., {Blake}, C., {et~al.} 2010, \mnras, 401,
  1429, \dodoi{10.1111/j.1365-2966.2009.15754.x}

\bibitem[{{Eriksen} {et~al.}(2008){Eriksen}, {Dickinson}, {Jewell}, {Banday},
  {G{\'o}rski}, \& {Lawrence}}]{2008ApJ...672L..87E}
{Eriksen}, H.~K., {Dickinson}, C., {Jewell}, J.~B., {et~al.} 2008, \apjl, 672,
  L87, \dodoi{10.1086/526545}

\bibitem[{Eriksen {et~al.}(2008)Eriksen, Jewell, Dickinson, Banday,
  G{\'{o}}rski, \& Lawrence}]{Eriksen_2008}
Eriksen, H.~K., Jewell, J.~B., Dickinson, C., {et~al.} 2008, The Astrophysical
  Journal, 676, 10, \dodoi{10.1086/525277}

\bibitem[{{Eriksen} {et~al.}(2008){Eriksen}, {Jewell}, {Dickinson}, {Banday},
  {G{\'o}rski}, \& {Lawrence}}]{2008ApJ...676...10E}
{Eriksen}, H.~K., {Jewell}, J.~B., {Dickinson}, C., {et~al.} 2008, \apj, 676,
  10, \dodoi{10.1086/525277}

\bibitem[{{Ewall-Wice} {et~al.}(2018){Ewall-Wice}, {Chang}, {Lazio},
  {Dor{\'e}}, {Seiffert}, \& {Monsalve}}]{2018ApJ...868...63E}
{Ewall-Wice}, A., {Chang}, T.~C., {Lazio}, J., {et~al.} 2018, \apj, 868, 63,
  \dodoi{10.3847/1538-4357/aae51d}

\bibitem[{{Ewall-Wice} {et~al.}(2021){Ewall-Wice}, {Kern}, {Dillon}, {Liu},
  {Parsons}, {Singh}, {Lanman}, {La Plante}, {Fagnoni}, {Acedo}, {DeBoer},
  {Nunhokee}, {Bull}, {Chang}, {Lazio}, {Aguirre}, \&
  {Weinberg}}]{2021MNRAS.500.5195E}
{Ewall-Wice}, A., {Kern}, N., {Dillon}, J.~S., {et~al.} 2021, \mnras, 500,
  5195, \dodoi{10.1093/mnras/staa3293}

\bibitem[{{Feng} \& {Holder}(2018)}]{2018ApJ...858L..17F}
{Feng}, C., \& {Holder}, G. 2018, \apjl, 858, L17,
  \dodoi{10.3847/2041-8213/aac0fe}

\bibitem[{{Fern{\'a}ndez-Cobos} {et~al.}(2012){Fern{\'a}ndez-Cobos}, {Vielva},
  {Barreiro}, \& {Mart{\'\i}nez-Gonz{\'a}lez}}]{2012MNRAS.420.2162F}
{Fern{\'a}ndez-Cobos}, R., {Vielva}, P., {Barreiro}, R.~B., \&
  {Mart{\'\i}nez-Gonz{\'a}lez}, E. 2012, \mnras, 420, 2162,
  \dodoi{10.1111/j.1365-2966.2011.20182.x}

\bibitem[{{Fialkov} \& {Barkana}(2019)}]{2019MNRAS.486.1763F}
{Fialkov}, A., \& {Barkana}, R. 2019, \mnras, 486, 1763,
  \dodoi{10.1093/mnras/stz873}

\bibitem[{{Furlanetto} {et~al.}(2006){Furlanetto}, {Oh}, \&
  {Briggs}}]{2006PhR...433..181F}
{Furlanetto}, S.~R., {Oh}, S.~P., \& {Briggs}, F.~H. 2006, \physrep, 433, 181,
  \dodoi{10.1016/j.physrep.2006.08.002}

\bibitem[{{Ghosh} {et~al.}(2015){Ghosh}, {Koopmans}, {Chapman}, \&
  {Jeli{\'c}}}]{2015MNRAS.452.1587G}
{Ghosh}, A., {Koopmans}, L. V.~E., {Chapman}, E., \& {Jeli{\'c}}, V. 2015,
  \mnras, 452, 1587, \dodoi{10.1093/mnras/stv1355}

\bibitem[{Gold {et~al.}(2009)Gold, Bennett, Hill, Hinshaw, Odegard, Page,
  Spergel, Weiland, Dunkley, Halpern, Jarosik, Kogut, Komatsu, Larson, Meyer,
  Nolta, Wollack, \& Wright}]{Gold_2009}
Gold, B., Bennett, C.~L., Hill, R.~S., {et~al.} 2009, The Astrophysical Journal
  Supplement Series, 180, 265, \dodoi{10.1088/0067-0049/180/2/265}

\bibitem[{{G{\'o}rski} {et~al.}(2005){G{\'o}rski}, {Hivon}, {Banday},
  {Wandelt}, {Hansen}, {Reinecke}, \& {Bartelmann}}]{2005ApJ...622..759G}
{G{\'o}rski}, K.~M., {Hivon}, E., {Banday}, A.~J., {et~al.} 2005, \apj, 622,
  759, \dodoi{10.1086/427976}

\bibitem[{{Hivon} {et~al.}(2002){Hivon}, {G{\'o}rski}, {Netterfield}, {Crill},
  {Prunet}, \& {Hansen}}]{2002ApJ...567....2H}
{Hivon}, E., {G{\'o}rski}, K.~M., {Netterfield}, C.~B., {et~al.} 2002, \apj,
  567, 2, \dodoi{10.1086/338126}

\bibitem[{{Joseph} {et~al.}(2023){Joseph}, {Purkayastha}, \&
  {Saha}}]{2023MNRAS.520..976J}
{Joseph}, A., {Purkayastha}, U., \& {Saha}, R. 2023, \mnras, 520, 976,
  \dodoi{10.1093/mnras/stad187}

\bibitem[{{Joseph} \& {Saha}(2021)}]{2021JApA...42..111J}
{Joseph}, A., \& {Saha}, R. 2021, Journal of Astrophysics and Astronomy, 42,
  111, \dodoi{10.1007/s12036-021-09776-6}

\bibitem[{{Joseph} \& {Saha}(2022)}]{2022MNRAS.511.1637J}
---. 2022, \mnras, 511, 1637, \dodoi{10.1093/mnras/stac201}

\bibitem[{{Joseph} \& {Saha}(2023)}]{2023MNRAS.519.1809J}
---. 2023, \mnras, 519, 1809, \dodoi{10.1093/mnras/stac3586}

\bibitem[{{Kaiser}(1987)}]{1987MNRAS.227....1K}
{Kaiser}, N. 1987, \mnras, 227, 1, \dodoi{10.1093/mnras/227.1.1}

\bibitem[{{Kennedy} {et~al.}(2023){Kennedy}, {Bull}, {Wilensky}, {Burba}, \&
  {Choudhuri}}]{2023ApJS..266...23K}
{Kennedy}, F., {Bull}, P., {Wilensky}, M.~J., {Burba}, J., \& {Choudhuri}, S.
  2023, \apjs, 266, 23, \dodoi{10.3847/1538-4365/acc324}

\bibitem[{{Kern} \& {Liu}(2021)}]{2021MNRAS.501.1463K}
{Kern}, N.~S., \& {Liu}, A. 2021, \mnras, 501, 1463,
  \dodoi{10.1093/mnras/staa3736}

\bibitem[{{Koopmans} {et~al.}(2015){Koopmans}, {Pritchard}, {Mellema},
  {Aguirre}, {Ahn}, {Barkana}, {van Bemmel}, {Bernardi}, {Bonaldi}, {Briggs},
  {de Bruyn}, {Chang}, {Chapman}, {Chen}, {Ciardi}, {Dayal}, {Ferrara},
  {Fialkov}, {Fiore}, {Ichiki}, {Illiev}, {Inoue}, {Jelic}, {Jones}, {Lazio},
  {Maio}, {Majumdar}, {Mack}, {Mesinger}, {Morales}, {Parsons}, {Pen},
  {Santos}, {Schneider}, {Semelin}, {de Souza}, {Subrahmanyan}, {Takeuchi},
  {Vedantham}, {Wagg}, {Webster}, {Wyithe}, {Datta}, \&
  {Trott}}]{2015aska.confE...1K}
{Koopmans}, L., {Pritchard}, J., {Mellema}, G., {et~al.} 2015, in Advancing
  Astrophysics with the Square Kilometre Array (AASKA14), 1,
  \dodoi{10.22323/1.215.0001}

\bibitem[{{Kovetz} {et~al.}(2019){Kovetz}, {Breysse}, {Lidz}, {Bock},
  {Bradford}, {Chang}, {Foreman}, {Padmanabhan}, {Pullen}, {Riechers}, {Silva},
  \& {Switzer}}]{2019BAAS...51c.101K}
{Kovetz}, E., {Breysse}, P.~C., {Lidz}, A., {et~al.} 2019, \baas, 51, 101,
  \dodoi{10.48550/arXiv.1903.04496}

\bibitem[{{Lah} {et~al.}(2007){Lah}, {Chengalur}, {Briggs}, {Colless}, {de
  Propris}, {Pracy}, {de Blok}, {Fujita}, {Ajiki}, {Shioya}, {Nagao},
  {Murayama}, {Taniguchi}, {Yagi}, \& {Okamura}}]{2007MNRAS.376.1357L}
{Lah}, P., {Chengalur}, J.~N., {Briggs}, F.~H., {et~al.} 2007, \mnras, 376,
  1357, \dodoi{10.1111/j.1365-2966.2007.11540.x}

\bibitem[{{Liu} \& {Shaw}(2020)}]{2020PASP..132f2001L}
{Liu}, A., \& {Shaw}, J.~R. 2020, \pasp, 132, 062001,
  \dodoi{10.1088/1538-3873/ab5bfd}

\bibitem[{{Liu} \& {Tegmark}(2011)}]{2011PhRvD..83j3006L}
{Liu}, A., \& {Tegmark}, M. 2011, \prd, 83, 103006,
  \dodoi{10.1103/PhysRevD.83.103006}

\bibitem[{{Loeb} \& {Zaldarriaga}(2004)}]{2004PhRvL..92u1301L}
{Loeb}, A., \& {Zaldarriaga}, M. 2004, \prl, 92, 211301,
  \dodoi{10.1103/PhysRevLett.92.211301}

\bibitem[{{Maino} {et~al.}(2003){Maino}, {Banday}, {Baccigalupi}, {Perrotta},
  \& {G{\'o}rski}}]{2003MNRAS.344..544M}
{Maino}, D., {Banday}, A.~J., {Baccigalupi}, C., {Perrotta}, F., \&
  {G{\'o}rski}, K.~M. 2003, \mnras, 344, 544,
  \dodoi{10.1046/j.1365-8711.2003.06835.x}

\bibitem[{{Maino} {et~al.}(2002){Maino}, {Farusi}, {Baccigalupi}, {Perrotta},
  {Banday}, {Bedini}, {Burigana}, {De Zotti}, {G{\'o}rski}, \&
  {Salerno}}]{2002MNRAS.334...53M}
{Maino}, D., {Farusi}, A., {Baccigalupi}, C., {et~al.} 2002, \mnras, 334, 53,
  \dodoi{10.1046/j.1365-8711.2002.05425.x}

\bibitem[{{Masui} {et~al.}(2013){Masui}, {Switzer}, {Banavar}, {Bandura},
  {Blake}, {Calin}, {Chang}, {Chen}, {Li}, {Liao}, {Natarajan}, {Pen},
  {Peterson}, {Shaw}, \& {Voytek}}]{2013ApJ...763L..20M}
{Masui}, K.~W., {Switzer}, E.~R., {Banavar}, N., {et~al.} 2013, \apjl, 763,
  L20, \dodoi{10.1088/2041-8205/763/1/L20}

\bibitem[{{McQuinn} {et~al.}(2006){McQuinn}, {Zahn}, {Zaldarriaga},
  {Hernquist}, \& {Furlanetto}}]{2006ApJ...653..815M}
{McQuinn}, M., {Zahn}, O., {Zaldarriaga}, M., {Hernquist}, L., \& {Furlanetto},
  S.~R. 2006, \apj, 653, 815, \dodoi{10.1086/505167}

\bibitem[{{Mertens} {et~al.}(2018){Mertens}, {Ghosh}, \&
  {Koopmans}}]{2018MNRAS.478.3640M}
{Mertens}, F.~G., {Ghosh}, A., \& {Koopmans}, L.~V.~E. 2018, \mnras, 478, 3640,
  \dodoi{10.1093/mnras/sty1207}

\bibitem[{{Miville-Desch{\^e}nes} {et~al.}(2008){Miville-Desch{\^e}nes},
  {Ysard}, {Lavabre}, {Ponthieu}, {Mac{\'\i}as-P{\'e}rez}, {Aumont}, \&
  {Bernard}}]{2008A&A...490.1093M}
{Miville-Desch{\^e}nes}, M.~A., {Ysard}, N., {Lavabre}, A., {et~al.} 2008,
  \aap, 490, 1093, \dodoi{10.1051/0004-6361:200809484}

\bibitem[{{Morales} \& {Wyithe}(2010)}]{2010ARA&A..48..127M}
{Morales}, M.~F., \& {Wyithe}, J. S.~B. 2010, \araa, 48, 127,
  \dodoi{10.1146/annurev-astro-081309-130936}

\bibitem[{{Mu{\~n}oz} \& {Loeb}(2018)}]{2018arXiv180210094M}
{Mu{\~n}oz}, J.~B., \& {Loeb}, A. 2018, arXiv e-prints, arXiv:1802.10094,
  \dodoi{10.48550/arXiv.1802.10094}

\bibitem[{{Olivari} {et~al.}(2016){Olivari}, {Remazeilles}, \&
  {Dickinson}}]{2016MNRAS.456.2749O}
{Olivari}, L.~C., {Remazeilles}, M., \& {Dickinson}, C. 2016, \mnras, 456,
  2749, \dodoi{10.1093/mnras/stv2884}

\bibitem[{{Paciga} {et~al.}(2013){Paciga}, {Albert}, {Bandura}, {Chang},
  {Gupta}, {Hirata}, {Odegova}, {Pen}, {Peterson}, {Roy}, {Shaw}, {Sigurdson},
  \& {Voytek}}]{2013MNRAS.433..639P}
{Paciga}, G., {Albert}, J.~G., {Bandura}, K., {et~al.} 2013, \mnras, 433, 639,
  \dodoi{10.1093/mnras/stt753}

\bibitem[{{Planck Collaboration}(2014)}]{2014A&A...571A..16P}
{Planck Collaboration}. 2014, \aap, 571, A16,
  \dodoi{10.1051/0004-6361/201321591}

\bibitem[{{Planck Collaboration}(2016)}]{2016A&A...594A..10P}
---. 2016, \aap, 594, A10, \dodoi{10.1051/0004-6361/201525967}

\bibitem[{{Planck Collaboration}(2020)}]{2020A&A...641A...6P}
---. 2020, \aap, 641, A6, \dodoi{10.1051/0004-6361/201833910}

\bibitem[{{Pritchard} \& {Loeb}(2012)}]{2012RPPh...75h6901P}
{Pritchard}, J.~R., \& {Loeb}, A. 2012, Reports on Progress in Physics, 75,
  086901, \dodoi{10.1088/0034-4885/75/8/086901}

\bibitem[{{Remazeilles} {et~al.}(2011){Remazeilles}, {Delabrouille}, \&
  {Cardoso}}]{2011MNRAS.418..467R}
{Remazeilles}, M., {Delabrouille}, J., \& {Cardoso}, J.-F. 2011, \mnras, 418,
  467, \dodoi{10.1111/j.1365-2966.2011.19497.x}

\bibitem[{{Remazeilles} {et~al.}(2015){Remazeilles}, {Dickinson}, {Banday},
  {Bigot-Sazy}, \& {Ghosh}}]{2015MNRAS.451.4311R}
{Remazeilles}, M., {Dickinson}, C., {Banday}, A.~J., {Bigot-Sazy}, M.~A., \&
  {Ghosh}, T. 2015, \mnras, 451, 4311, \dodoi{10.1093/mnras/stv1274}

\bibitem[{{Saha} {et~al.}(2008){Saha}, {Prunet}, {Jain}, \&
  {Souradeep}}]{2008PhRvD..78b3003S}
{Saha}, R., {Prunet}, S., {Jain}, P., \& {Souradeep}, T. 2008, \prd, 78,
  023003, \dodoi{10.1103/PhysRevD.78.023003}

\bibitem[{{Scott} \& {Rees}(1990)}]{1990MNRAS.247..510S}
{Scott}, D., \& {Rees}, M.~J. 1990, \mnras, 247, 510.
\newblock \url{https://ui.adsabs.harvard.edu/abs/1990MNRAS.247..510S}

\bibitem[{{Shaw} {et~al.}(2014){Shaw}, {Sigurdson}, {Pen}, {Stebbins}, \&
  {Sitwell}}]{2014ApJ...781...57S}
{Shaw}, J.~R., {Sigurdson}, K., {Pen}, U.-L., {Stebbins}, A., \& {Sitwell}, M.
  2014, \apj, 781, 57, \dodoi{10.1088/0004-637X/781/2/57}

\bibitem[{{Shaw} {et~al.}(2015){Shaw}, {Sigurdson}, {Sitwell}, {Stebbins}, \&
  {Pen}}]{2015PhRvD..91h3514S}
{Shaw}, J.~R., {Sigurdson}, K., {Sitwell}, M., {Stebbins}, A., \& {Pen}, U.-L.
  2015, \prd, 91, 083514, \dodoi{10.1103/PhysRevD.91.083514}

\bibitem[{Souradeep {et~al.}(2006)Souradeep, Saha, \& Jain}]{SOURADEEP2006854}
Souradeep, T., Saha, R., \& Jain, P. 2006, New Astronomy Reviews, 50, 854,
  \dodoi{https://doi.org/10.1016/j.newar.2006.09.009}

\bibitem[{{Spinelli} {et~al.}(2022){Spinelli}, {Carucci}, {Cunnington},
  {Harper}, {Irfan}, {Fonseca}, {Pourtsidou}, \& {Wolz}}]{2022MNRAS.509.2048S}
{Spinelli}, M., {Carucci}, I.~P., {Cunnington}, S., {et~al.} 2022, \mnras, 509,
  2048, \dodoi{10.1093/mnras/stab3064}

\bibitem[{{Sudevan} \& {Saha}(2018)}]{2018ApJ...867...74S}
{Sudevan}, V., \& {Saha}, R. 2018, \apj, 867, 74,
  \dodoi{10.3847/1538-4357/aae439}

\bibitem[{{Switzer} {et~al.}(2013){Switzer}, {Masui}, {Bandura}, {Calin},
  {Chang}, {Chen}, {Li}, {Liao}, {Natarajan}, {Pen}, {Peterson}, {Shaw}, \&
  {Voytek}}]{2013MNRAS.434L..46S}
{Switzer}, E.~R., {Masui}, K.~W., {Bandura}, K., {et~al.} 2013, \mnras, 434,
  L46, \dodoi{10.1093/mnrasl/slt074}

\bibitem[{Tegmark {et~al.}(2003)Tegmark, de~Oliveira-Costa, \&
  Hamilton}]{PhysRevD.68.123523}
Tegmark, M., de~Oliveira-Costa, A., \& Hamilton, A. J.~S. 2003, Phys. Rev. D,
  68, 123523, \dodoi{10.1103/PhysRevD.68.123523}

\bibitem[{Tegmark \& Efstathiou(1996)}]{10.1093/mnras/281.4.1297}
Tegmark, M., \& Efstathiou, G. 1996, Monthly Notices of the Royal Astronomical
  Society, 281, 1297, \dodoi{10.1093/mnras/281.4.1297}

\bibitem[{{Thorne} {et~al.}(2017){Thorne}, {Dunkley}, {Alonso}, \&
  {N{\ae}ss}}]{2017MNRAS.469.2821T}
{Thorne}, B., {Dunkley}, J., {Alonso}, D., \& {N{\ae}ss}, S. 2017, \mnras, 469,
  2821, \dodoi{10.1093/mnras/stx949}

\bibitem[{{Villaescusa-Navarro} {et~al.}(2018){Villaescusa-Navarro}, {Genel},
  {Castorina}, {Obuljen}, {Spergel}, {Hernquist}, {Nelson}, {Carucci},
  {Pillepich}, {Marinacci}, {Diemer}, {Vogelsberger}, {Weinberger}, \&
  {Pakmor}}]{2018ApJ...866..135V}
{Villaescusa-Navarro}, F., {Genel}, S., {Castorina}, E., {et~al.} 2018, \apj,
  866, 135, \dodoi{10.3847/1538-4357/aadba0}

\bibitem[{{Wandelt} {et~al.}(2004){Wandelt}, {Larson}, \&
  {Lakshminarayanan}}]{2004PhRvD..70h3511W}
{Wandelt}, B.~D., {Larson}, D.~L., \& {Lakshminarayanan}, A. 2004, \prd, 70,
  083511, \dodoi{10.1103/PhysRevD.70.083511}

\bibitem[{{Wang} {et~al.}(2024){Wang}, {Masui}, {Bandura}, {Chakraborty},
  {Dobbs}, {Foreman}, {Gray}, {Halpern}, {Joseph}, {MacEachern}, {Mena-Parra},
  {Miller}, {Newburgh}, {Paul}, {Reda}, {Sanghavi}, {Siegel}, \&
  {Wulf}}]{2024arXiv240808949W}
{Wang}, H., {Masui}, K., {Bandura}, K., {et~al.} 2024, arXiv e-prints,
  arXiv:2408.08949, \dodoi{10.48550/arXiv.2408.08949}

\bibitem[{{Wang} {et~al.}(2021){Wang}, {Santos}, {Bull}, {Grainge},
  {Cunnington}, {Fonseca}, {Irfan}, {Li}, {Pourtsidou}, {Soares}, {Spinelli},
  {Bernardi}, \& {Engelbrecht}}]{2021MNRAS.505.3698W}
{Wang}, J., {Santos}, M.~G., {Bull}, P., {et~al.} 2021, \mnras, 505, 3698,
  \dodoi{10.1093/mnras/stab1365}

\bibitem[{{White} {et~al.}(1999){White}, {Carlstrom}, {Dragovan}, \&
  {Holzapfel}}]{1999ApJ...514...12W}
{White}, M., {Carlstrom}, J.~E., {Dragovan}, M., \& {Holzapfel}, W.~L. 1999,
  \apj, 514, 12, \dodoi{10.1086/306911}

\bibitem[{{Wilson} {et~al.}(2013){Wilson}, {Rohlfs}, \&
  {H{\"u}ttemeister}}]{2013tra..book.....W}
{Wilson}, T.~L., {Rohlfs}, K., \& {H{\"u}ttemeister}, S. 2013, {Tools of Radio
  Astronomy}, \dodoi{10.1007/978-3-642-39950-3}

\bibitem[{{Wolz} {et~al.}(2017){Wolz}, {Blake}, {Abdalla}, {Anderson}, {Chang},
  {Li}, {Masui}, {Switzer}, {Pen}, {Voytek}, \& {Yadav}}]{2017MNRAS.464.4938W}
{Wolz}, L., {Blake}, C., {Abdalla}, F.~B., {et~al.} 2017, \mnras, 464, 4938,
  \dodoi{10.1093/mnras/stw2556}

\bibitem[{{Zhang} {et~al.}(2016){Zhang}, {Bunn}, {Karakci}, {Korotkov},
  {Sutter}, {Timbie}, {Tucker}, \& {Wandelt}}]{2016ApJS..222....3Z}
{Zhang}, L., {Bunn}, E.~F., {Karakci}, A., {et~al.} 2016, \apjs, 222, 3,
  \dodoi{10.3847/0067-0049/222/1/3}

\bibitem[{{Zuo} {et~al.}(2019){Zuo}, {Chen}, {Ansari}, \&
  {Lu}}]{2019AJ....157....4Z}
{Zuo}, S., {Chen}, X., {Ansari}, R., \& {Lu}, Y. 2019, \aj, 157, 4,
  \dodoi{10.3847/1538-3881/aaef3b}

\end{thebibliography}
\bibliographystyle{aasjournal}



\end{document}